\documentclass[fleqn,usenatbib]{mnras}
\pdfoutput=1
\usepackage{newtxtext,newtxmath}
\usepackage{xcolor}
\usepackage{multirow}
\usepackage{threeparttable}
\usepackage{subcaption}
\usepackage{mwe}
\usepackage{comment}

\usepackage[T1]{fontenc}

\DeclareRobustCommand{\VAN}[3]{#2}
\let\VANthebibliography\thebibliography
\def\thebibliography{\DeclareRobustCommand{\VAN}[3]{##3}\VANthebibliography}

\usepackage{graphicx,caption}	
\usepackage{amsmath}	

\usepackage{footnote}
\makesavenoteenv{table}
\makesavenoteenv{tabular}
\usepackage{graphicx}
\usepackage{soul}
\usepackage{hyperref}

\title[Subpulse Drifting in J0026--1955]{PSR J0026--1955: A curious case of evolutionary subpulse drifting and nulling}

\author[P. Janagal et al.]{Parul Janagal,$^{1}$\thanks{E-mail: phd1801121004@iiti.ac.in}
Samuel J. McSweeney$^{2}$,
Manoneeta Chakraborty,$^{1}$
N. D. Ramesh Bhat$^{2}$ 
\\
$^{1}$Department of Astronomy, Astrophysics, and Space Engineering, Indian Institute of Technology Indore, Indore 453552, India\\
$^{2}$International Centre for Radio Astronomy Research, Curtin University, Bentley, WA 6102, Australia\\
}

\date{Accepted XXX. Received YYY; in original form ZZZ}

\pubyear{2023}

\begin{document}
\label{firstpage}
\pagerange{\pageref{firstpage}--\pageref{lastpage}}
\maketitle

\begin{abstract}
PSR J0026–1955 was independently discovered by the Murchison Widefield Array (MWA) recently. The pulsar exhibits subpulse drifting, where the radio emission from a pulsar appears to drift in spin phase within the main pulse profile, and nulling, where the emission ceases briefly. The pulsar showcases a curious case of drift rate evolution as it exhibits rapid changes between the drift modes and a gradual evolution in the drift rate within a mode. Here we report new analysis and results from observations of J0026–1955 made with the upgraded Giant Meterwave Radio Telescope (uGMRT) at 300-500 MHz. We identify two distinct subpulse drifting modes: A and B, with mode A sub-categorised into A0, A1, and A2, depending upon the drift rate evolutionary behaviour. Additionally, the pulsar exhibits short and long nulls, with an estimated overall nulling fraction of $\sim$58\%, which is lower than the previously reported value. Our results also provide evidence of subpulse memory across nulls and a consistent behaviour where mode A2 is often followed by a null. We investigate the drift rate modulations of J0026–1955 and put forward two different models to explain the observed drifting behaviour. We suggest that either a change in polar gap screening or a slow relaxation in the spark configuration could possibly drive the evolution in drift rates. J0026–1955 belongs to a rare subset of pulsars which exhibit subpulse drifting, nulling, mode changing, and drift rate evolution. It is, therefore, an ideal test bed for carousel models and to uncover the intricacies of pulsar emission physics.

\end{abstract}

\begin{keywords}
stars: neutron - pulsars: general - pulsars: individual (PSR J0026-1955)
\end{keywords}



\section{Introduction}

Radio pulsars are rotating neutron stars with highly coherent radiation emanating from the vicinity of magnetic poles, which cross our line of sight once every pulsar rotation \citep{1968Natur.217..709H}. They possess a large mass ($\sim$1 to $\sim$2 $M_{\odot}$) confined within a small radius ($\lesssim$ 10 km), with strong gravitational ($\sim10^{11}$ times stronger than the Earth's surface gravitational field) and magnetic fields \cite[$\sim 10^{8}$ to $\sim 10^{12}$ G; e.g.,][]{1969Natur.221...25G, 2004hpa..book.....L}. Pulsars are the sites of some of the highest energy physical processes, making them powerful astrophysical laboratories, owing to such extreme environments of very strong gravitational and magnetic fields surrounding them. However, even though more than 3000 pulsars are known to date, a definitive exposition of the processes by which pulsars emit beams of radio waves is still non-existent in the literature \citep[e.g.,][]{10.1093/mnras/staa3324}.

Radio emission from pulsars exhibits a variety of phenomena, which modulate their pulse-to-pulse emission, observable in the form of subpulse drifting, nulling, mode changing, etc., thereby providing a range of avenues to understand the complex physical processes that cause the emission. In many cases, individual pulses from a pulsar show substructure with one or more distinct components called subpulses. \cite{1968Natur.220..231D} observed the systematic `marching' of these subpulses with phase within the on-pulse window, leading to diagonal drifting structures in a pulse stack, called driftbands (pulse number vs rotation phase), as commonly seen in many pulsars \citep[e.g.,][]{10.1111/j.1365-2966.2009.15210.x, 2017ApJ...836..224M,2018A&A...616A.119B, 2022MNRAS.509.4573J}. For such a pulse stack, the drift rate is then defined as the reciprocal of the slope of the driftbands ($^\circ/P_1$, where $P_1$ is the pulsar rotation period).

Theoretical models explaining subpulse drifting were suggested early on after the discovery of subpulse drifting in pulsars \citep[e.g.,][]{1971ApJ...164..529S,1975ApJ...196...51R}. The most well-developed model at the time was able to explain the subpulse drifting phenomenon exhibited by pulsars studied then, most with stable drift rates, such as B0809+74 and B0943+10 \citep[e.g.,][]{1971ApL.....9..205T, 1999ApJ...524.1008D}. The original model proposed by \cite{1975ApJ...196...51R} associated drifting subpulses with a rotating `carousel' of a discrete number of sparks (electrical discharges) in regions of charge depletion just above the neutron star surface near the magnetic poles. This carousel of sparks circulates around the magnetic axis due to an \textbf{E} $\times$ \textbf{B} drift, and the electron-positron pairs produced in the discharges are ultimately responsible for the observed radio emission. The rotation rate of the carousel ($P_4$) around the magnetic axis is generally different from the pulsar period. 

Two characteristic features of subpulse drifting pulsars are their drift rates and $P_3$. The drift rate is defined as $D = \Delta \phi$ per pulse period ($^{\circ}/P_1$), where $\Delta \phi$ is the longitude shift in degrees during one pulse period $P_1$. A positive value indicates a drift from early to later longitudes, while a negative value corresponds to a drift from later to earlier longitudes. In a pulse stack, the vertical separation between driftbands at a given longitude is $ P_3 $ (typically expressed in units of the pulsar rotation period, $ P_1 $), which is a measure of time after which a subpulse will return at a particular phase. The caveat here is that the pulsar rotation only permits observation of the subpulse positions once every pulse. A specific subpulse in one pulse cannot be unambiguously identified in the next due to the difficulty in resolving the presence of aliasing, making it generally difficult to evaluate the true carousel speed. That is to say, if aliasing is present, the observed drift rate is related to the beating frequency between $P_1$ and $P_4$. Another consequence of aliasing is that the drift rate can appear to vary even if $P_4$ stays constant as long as the beamlet configuration changes. Thus, if multiple drift rates are present in a given pulsar, it does not necessarily mean that the rotation speed of the carousel has changed; it may be that the number of beamlets has changed instead \citep[e.g.,][]{10.1093/mnras/stt739, 2017ApJ...836..224M, 2022MNRAS.509.4573J}.

Even in the simplest case of a constant $P_4$ and a fixed number of beamlets, the apparent drift rate (i.e. the slope of the driftbands) is not a steady function of rotation longitude. This is a purely geometric phenomenon related to the projection of the beamlets' motion onto the line of sight trajectory, as explained in \cite{2002A&A...393..733E}. This results in the driftbands themselves appearing curved, referred to as ``geometric curvature''. Geometric curvature is always present but, similar to the polarisation position angle (PPA) of the rotating vector model \citep{1969ApL.....3..225R}, will only be visible if the pulse window is sufficiently wide for a given pulsar's particular viewing geometry. Geometric curvature is also similar to the PPA in that it is symmetric about the fiducial point, giving the driftbands a characteristic `S'-shape, with an excess (or deficit) of the drift rate appearing in the peripheral part of the pulse window.

The carousel model satisfactorily explains the subpulse drifting nature of some pulsars that show stable drift rates, citing the theoretical stability of electric and magnetic fields at the spark locations \citep{1975ApJ...196...51R}. Furthermore, multi-frequency observations of a large fraction of the pulsar population have brought forth a variety of such atypical pulsars \citep[e.g.,][]{2007A&A...469..607W, 2023MNRAS.520.4562S}. These studies show that a substantial fraction ($\sim$50\%) of known pulsars exhibit subpulse drifting. However, explaining the drifting behaviour in pulsars that exhibit anything more complicated than a single stable drift rate requires modifications or extensions to the basic carousel model. Such pulsars present ideal test beds to modify the classical carousel model. Several extensions have been proposed over the years to account for the observed complicated behaviour. For example, \cite{2000ApJ...541..351G} suggest that a quasi-central spark can account for the non-drifting core components in profiles. The well-known phenomenon of bi-drifting may be explained in terms of the presence of an inner annular gap \citep{Qiao_2004} or an inner acceleration gap \citep{2019MNRAS.482.3757B}, or non-circular spark motions \citep{2017MNRAS.464.2597W}. Similarly, the phenomenon of drift rate reversal, shown by some pulsars, can be explained by the modified carousel model, where sparks rotate around the location of the electric potential extremum of the polar cap instead of the magnetic axis \citep{2022ApJ...934...23S}. These extensions/modifications are generally developed to explain specific drifting behaviours observed in a relatively small subset of subpulse drifting pulsars. However, there is still no comprehensive theory that can describe all the observed drifting behaviours. 

Several theories have also been suggested to interpret the drifting subpulses geometrically. \cite{1991AuJPh..44..573K} and \cite{2005MNRAS.360..669G} suggested that drifting subpulses result from modulation in the emission region caused by drift waves in some form of magnetospheric oscillations. In their model, the subpulses result from periodic variations in the magnetospheric plasma, which may cause the emission region to move across the observer's line of sight. \cite{2008ApJ...680..664C} suggest that non-radial oscillations in the emission region could be responsible for subpulse drifting without invoking circulations in the magnetosphere. They propose that the drifting subpulses could be due to non-radial oscillations in the magnetosphere. Although these models are able to explain phenomena such as mode changing, other phenomena, such as bi-drifting, memory across nulls, etc., cannot be convincingly accounted for.

Another phenomenon often seen in conjunction with subpulse drifting is `nulling', where the emission from a pulsar ceases abruptly for a few to hundreds of pulse periods \citep{1970Natur.228.1297B} before it is restored. To date, pulse nulling has been reported in more than 200 pulsars \citep{2020A&A...644A..73W}, which is less than 10\% of the known pulsar population. Nulls lasting for one or two pulses are generally attributed to the stochastic processes within the pulsar magnetosphere \citep[e.g.,][]{Basu_2018}. However, in subpulse drifting pulsars, short nulls can be attributed to a slight variation of the spark distribution, where nulls are caused by an empty line-of-sight \citep[e.g.,][]{2022MNRAS.509.4573J}. Long nulls, on the other hand, are thought to be related to changes in the plasma processes within the pulsar magnetosphere \citep[e.g., PSR B1706-16;][]{10.1093/mnras/stx3284}, or even the spin-down energy loss in the most extreme cases \citep[e.g., PSR B1931+24;][]{2006Sci...312..549K}. If nulls and changes in the drift modes are, in fact, caused by intrinsic changes in the pulsar magnetosphere, their interactions could be crucial in understanding the mechanisms behind changes between different magnetospheric states. Nulling itself may be an extreme form of mode-changing, where a pulsar switches between different magnetospheric states, as suggested by the broadband behaviour of three nulling pulsars reported by \cite{10.1093/mnras/stt2389}. Hence, the study of pulsars exhibiting both nulling and drifting phenomena is crucial for comprehending the true origin and nature of the nulling phenomenon.

Pulsars which exhibit complicated drifting behaviour such as mode changing and nulling, and are also bright enough for single pulse analysis, are relatively rare. However, this combination is essential in shaping ideas concerning the pulsar radio emission process. Recently, the Murchison Widefield Array (MWA) independently discovered PSR J0026--1955 \citep{2022ApJ...933..210M} in the shallow pass of their Southern-Sky MWA Rapid Two-metre (SMART) pulsar survey \citep{smart1, smart2}. The pulsar was originally detected in 2018 in the Green Bank Northern Celestial Cap (GBNCC) pulsar survey \citep{2014ApJ...791...67S}. However, the discovery was not followed up until the recent re-discovery by the MWA. J0026--1955 is a bright pulsar which has a period of 1.306150 s and a dispersion measure (DM) of 20.869 pc cm$^{-3}$. The pulsar exhibits complex subpulse drifting behaviour and mode switching in addition to a large nulling fraction ($\sim$77\% at 155 MHz). \cite{2022ApJ...933..210M} found two distinct subpulse drifting modes A and B, with slow and fast drift rates, respectively, which were further categorised (A1/A2 and B1/B2) depending upon the qualitative properties of modal appearances and context. The pulsar was sometimes seen to abruptly change its drift rate, while at other times, it exhibits a consistent evolution of the drift rate within its drifting modes. 

The most distinctive feature of PSR J0026--1955 is its slow drift rate evolution, which has been found in only a handful of other pulsars like B0031-07 \citep{Vivekanand_1997,2000MNRAS.316..716J,2017ApJ...836..224M} and B0818-13 \citep{10.1093/mnras/204.2.519}. Furthermore, with its variable drift rates, the pulsar also poses an essential question to the stability of the carousel, as the basic models assume a stable configuration, leading to a non-variable drift rate throughout a drift mode. For J0026--1955, \cite{2022ApJ...933..210M} attempt to model the observed drifting behaviour with an exponentially decaying drift rate, similar to what is seen in PSR B0818-13 and PSR B0809+74 \citep{10.1093/mnras/204.2.519}. However, they were unable to fully characterise the observed drifting behaviour owing to their limited data sets and, consequently, an insufficient number of drift sequences. They also suggest a possibility of subpulse phase memory across short null sequences, which would benefit from more observations. Using longer multi-frequency observations, the modal taxonomy presented by \cite{2022ApJ...933..210M} can be tested for viability and to examine whether there is a need for something more sophisticated than an exponential model of drift rates. The complex drifting and nulling behaviour of this new pulsar thus warrants deeper investigations of its properties and their nature at different frequencies.

The unusual behaviour of J0026--1955 reported in \cite{2022ApJ...933..210M} will also benefit from a detailed study at higher frequencies, allowing us to test the frequency dependence of such characteristics. Furthermore, given the slow period and large nulling fraction of the pulsar, long-duration observations with a higher signal-to-noise ratio (S/N) are necessary to collect a sufficiently large number of complete burst sequences in order to undertake a more robust statistical analysis. 

In this study, we present a detailed investigation of subpulse drifting and nulling exhibited by J0026--1955, with new observations obtained using the upgraded Giant Metrewave Radio Telescope (uGMRT) at 300-500 MHz. This paper is organised as follows. In section \ref{sec:observations}, we briefly describe the observation details and data-reduction procedures; the subpulse drifting and nulling analysis are presented in section \ref{sec:analysis}; our findings are discussed in section \ref{sec:discussion}; and a summary of the paper is given in section \ref{sec:conclusion}.
\section{Observations and Data Reduction} \label{sec:observations}

The Giant Metrewave Radio Telescope (GMRT) is a radio interferometric array consisting of 30 antennas, each with a 45-meter diameter, and spread over an area of 28 square kilometres in a Y-shape \citep{1991CuSc...60...95S}. The GMRT recently underwent an upgrade, which included the addition of wide-band receivers and digital instrumentation, allowing for near-seamless coverage in frequency from 120 MHz to 1600 MHz \citep{2017CSci..113..707G, 2017JAI.....641011R}. For our observations of PSR J0026--1955, we used the upgraded GMRT in the phased array mode, where signals from each antenna are coherently added for maximum sensitivity. J0026--1955 was observed with the uGMRT at Band 3 (300-500 MHz) and Band 4 (550-750 MHz), over two epochs at each frequency band. However, due to the presence of higher levels of radio frequency interference (RFI), Band 4 data quality was not adequate for meaningful single-pulse analysis. Thus for the work presented in this paper, we limit our analysis to Band 3 (300-500 MHz) data. Details of observations, including the number of pulses which had pulsar emission (burst) and lacked any emission (null), are summarised in Table \ref{tab:obsdetails}. 

\begin{table}
\begin{center}
\begin{tabular}{|c|c|c|cc|}
\hline
\textbf{MJD}       & \textbf{Scan}      & \textbf{Length}             & \multicolumn{2}{c|}{\textbf{Number of Pulses}}               \\ \hline
\textbf{} & \textbf{} & \textit{(minutes)} & \multicolumn{1}{c|}{\textit{Burst}} & \textit{Null} \\ \hline
59529     & 1         & 54                 & \multicolumn{1}{c|}{1825}           & 669           \\ 
          & 2         & 43                 & \multicolumn{1}{c|}{735}            & 1234          \\ 
          & 3         & 54                 & \multicolumn{1}{c|}{781}            & 1717          \\ 
59543    & 1         & 54                 & \multicolumn{1}{c|}{971}            & 1506          \\ 
          & 2         & 54                 & \multicolumn{1}{c|}{642}            & 1841          \\ 
          & 3         & 54                 & \multicolumn{1}{c|}{1315}           & 1151          \\ \hline
Total     &           &                    & \multicolumn{1}{c|}{6269}           & 8118          \\ \hline
\end{tabular}
\begin{tablenotes}
    \small
    \item All observations are in the 300-500 MHz band, with 655.56$\mu$s time resolution, spread over 2048 channels.
\end{tablenotes}
\caption{Table of observations for PSR J0026--1955.}  
\label{tab:obsdetails}
\end{center}
\end{table}

The data were recorded at 655.56$\mu$s time resolution, spread across 2048 channels, thus providing a frequency resolution of 97.65 kHz, and were converted to single-pulse archives using the \texttt{DSPSR} package \citep{2011PASA...28....1V}. The single pulse files were then frequency scrunched and combined using the routines from \texttt{PSRCHIVE} \citep{2004PASA...21..302H}. Finally, each frequency-scrunched single-pulse sequence file was manually searched for RFI using the interactive RFI zapping subroutine \texttt{pazi} of \texttt{PSRCHIVE}. The RFI-excised file was then converted into an ASCII format that contained the pulse time series and was used for all subsequent analyses.

\begin{figure*}
     \centering
     \begin{subfigure}[b]{0.23\textwidth}
         \centering
         \includegraphics[width=\textwidth]{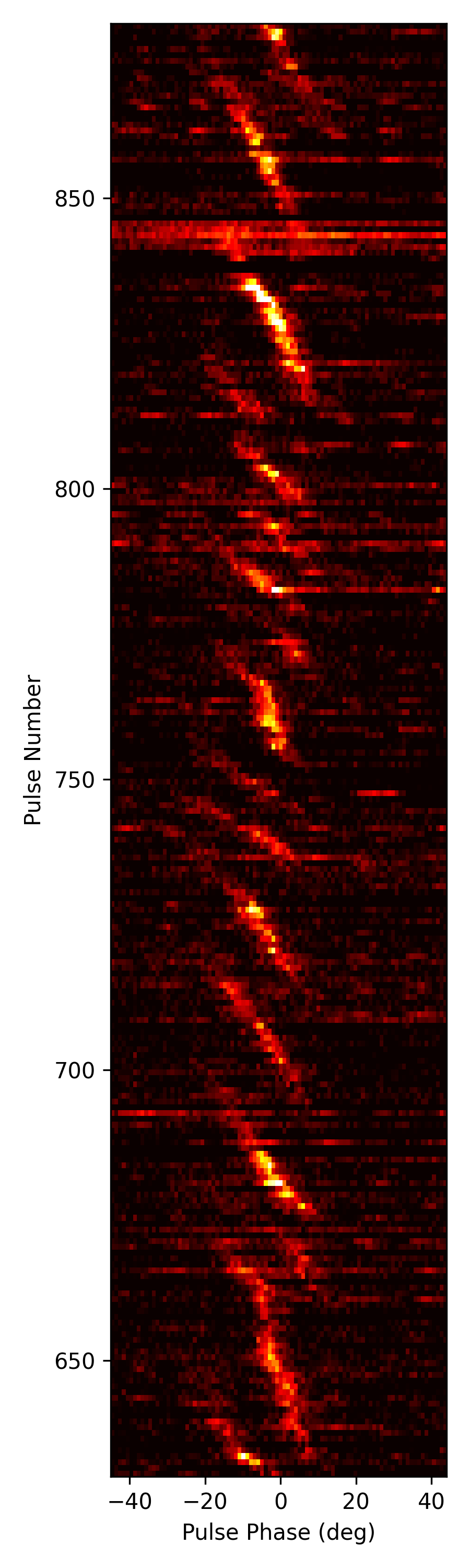}
         \caption{}
         \label{fig:fig1A}
     \end{subfigure}
     \begin{subfigure}[b]{0.23\textwidth}
         \centering
         \includegraphics[width=\textwidth]{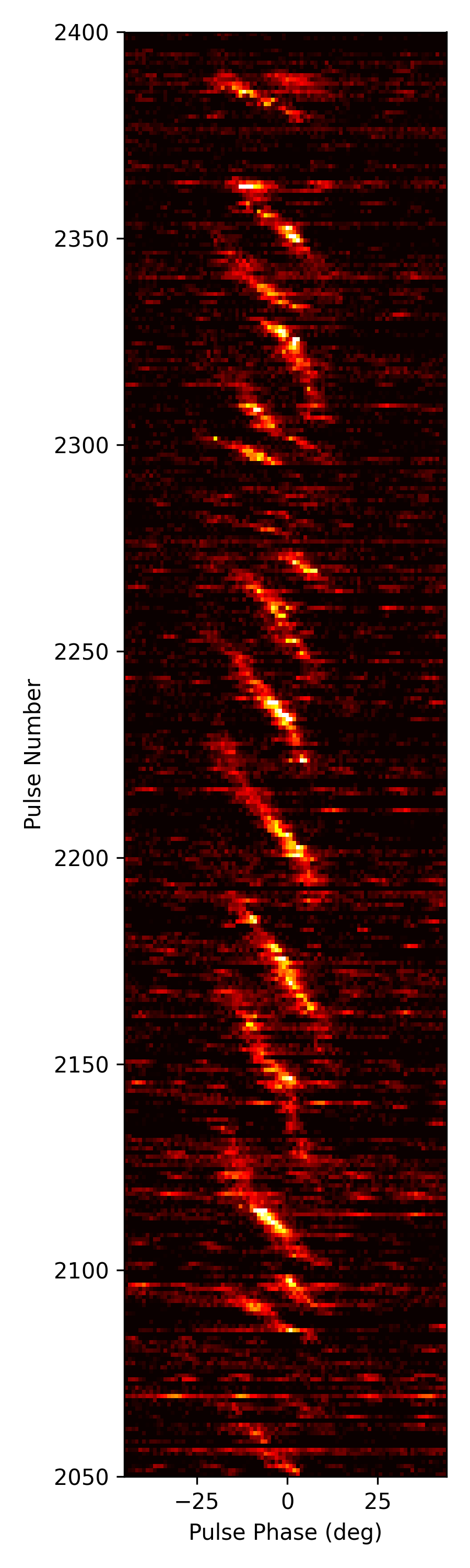}
         \caption{}
         \label{fig:fig1B}
     \end{subfigure}
     \begin{subfigure}[b]{0.23\textwidth}
         \centering
         \includegraphics[width=\textwidth]{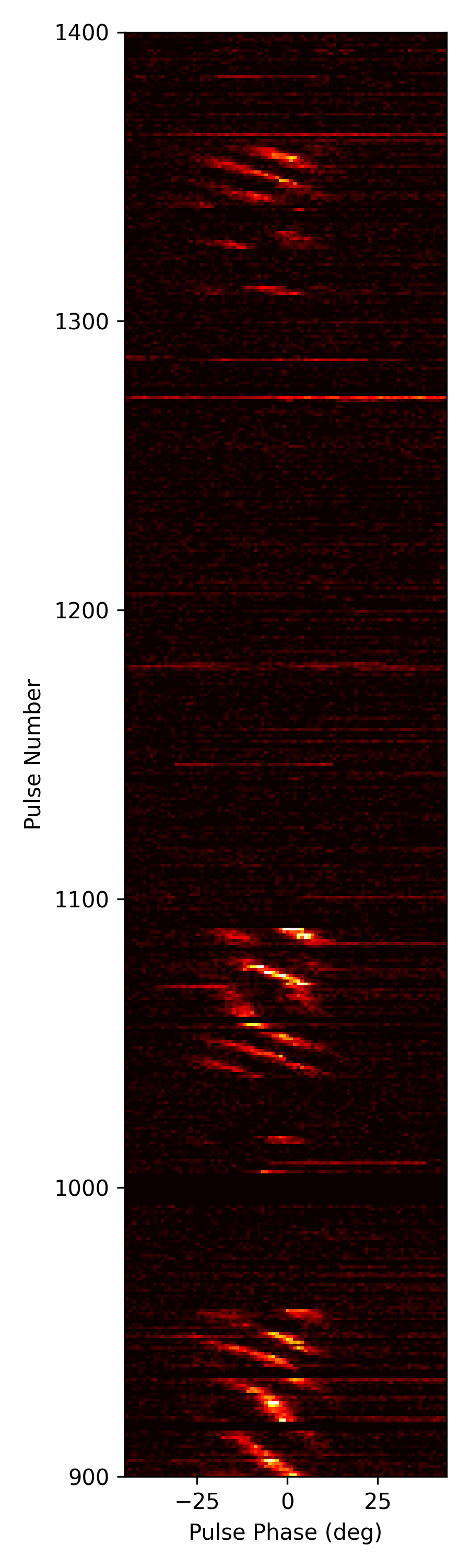}
         \caption{}
         \label{fig:fig1C}
     \end{subfigure}
     \begin{subfigure}[b]{0.23\textwidth}
         \centering
         \includegraphics[width=\textwidth]{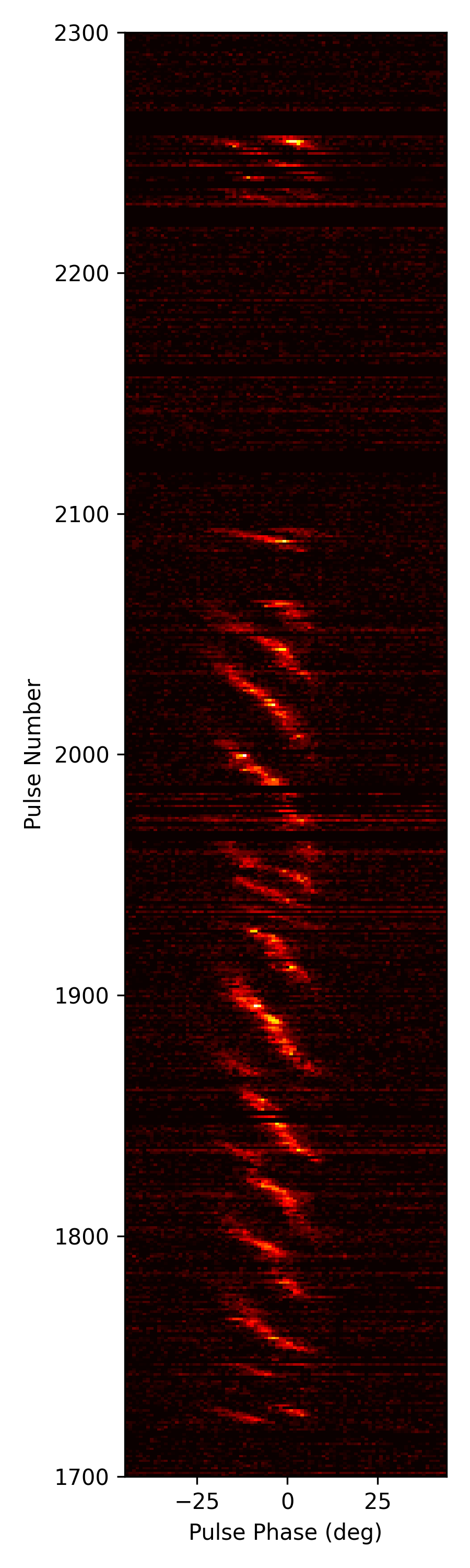}
         \caption{}
         \label{fig:fig1D}
     \end{subfigure}
        \caption{Each  pulse stack (Pulse phase vs Pulse number) represents the variety of subpulse drifting sequences shown by PSR J0026--1955 at 400 MHz (using uGMRT). Bright yellow diagonally arranged pixels around to phase $0^{\circ}$ are the driftbands. The pulsar exhibits both short and long nulls, where no emission can be seen in the on-pulse region.}
        \label{fig:figure1}
\end{figure*}

Following the methodology detailed above, the pulsar time series data obtained in the last step were used to generate the pulse stacks (pulse phase vs pulse number) as shown in Fig. \ref{fig:figure1}. The bright yellow diagonally arranged pixels between pulse phase $-20^{\circ}$ and $+20^{\circ}$ are the subpulse driftbands, which can be clearly seen to march from a positive to a negative phase, with increasing pulse number. Despite multiple rounds of RFI excision using the \texttt{pazi} subroutine, it is evident from Fig. \ref{fig:figure1} that there is some residual RFI. For example, several seconds of RFI can be seen right before pulse number 850 in panel \ref{fig:fig1A}. However, cases where the subpulses are bright enough to be visually recognised (despite the RFI), were retained. With such a tradeoff, we were able to salvage data that could still be used for exploration without affecting the subpulse drifting analysis. Further, panels \ref{fig:fig1C} and \ref{fig:fig1D}, provide clear examples of short and long nulls, where any emission from the pulsar is absent for a certain duration ranging from a few to a few hundred pulses. Table \ref{tab:obsdetails} presents a comprehensive summary of nulls and bursts observed in different observations.

\section{Analysis} \label{sec:analysis}

In subsequent analysis, we elaborate on different subpulse behaviours and attempt to characterise the drifting nature to learn the underlying mechanism in the context of the carousel model \citep{1975ApJ...196...51R}. 

In general, the preliminary analysis of any subpulse drifting pulsar aims at the classification of drift modes. However, considering that J0026--1955 does not always exhibit well-defined discrete modes but rather a drift rate evolution, such an analysis is complicated. In section \ref{bound}, we discuss a drift rate evolution-based classification scheme for deciding the mode boundaries, using linear and exponential models for drift rate evolution. Section \ref{modeclass} discusses the modes identified using this scheme. The observed nulling behaviour of the pulsar at 400 MHz is discussed in section \ref{null}. Further, the evolutionary drift rate behaviour of the pulsar is studied in detail in section \ref{drevol}. The exponential drift rate model used in section \ref{bound} was employed by \cite{10.1093/mnras/204.2.519} to demonstrate memory across nulls for the first time. Following their lead, we have also examined J0026--1955 for possible instances of memory across nulls, detailed in section \ref{sec:man}.

\subsection{Drift Mode Boundaries}\label{bound}

\begin{figure*}
    \centering
    \includegraphics[width=\textwidth]{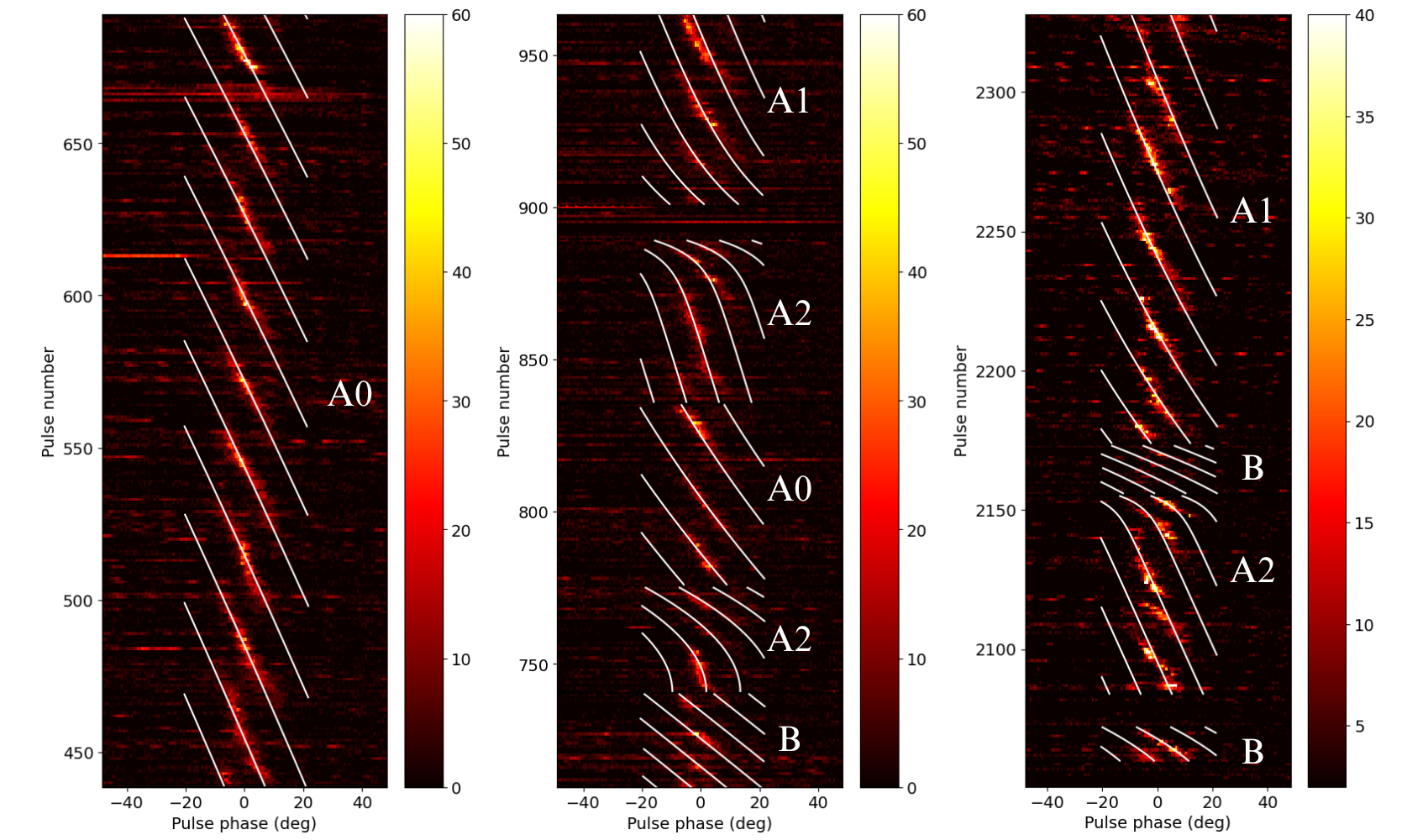}
    \caption{Pulse stack with driftbands (\textit{yellow bright subpulses}) modelled using linear and exponential models (\textit{white lines}). The evolution of drift rates can be seen in modes A1 and A2. The inter-modal driftband connectivity is also apparent in most cases.}
    \label{fig:figure2}
\end{figure*}

PSR J0026--1955 poses a unique challenge for identifying the drift modes. Generally, subpulse-drifting pulsars exhibit stable modes, which can be uniquely characterised by their $P_3$ values. However, the slowly evolving drift rates and $P_3$ of J0026--1955 complicate the mode identification. Thus, to identify the drift modes, we employed a different strategy. Given that the pulsar exhibits both evolutionary and non-evolutionary drift rates, we modelled the drift rate behaviours to make a drift rate-based modal classification, which also accounts for the evolution. The linear model is the most straightforward generalisation of a constant drift rate, essentially including the next term in the Taylor expansion -- and valid in situations where the drift sequences are sufficiently short relative to the rate of evolution.. Further, to account for the drift rate evolution across individual driftbands (which can show significant curvature), we follow the lead of \cite{10.1093/mnras/204.2.519} to use an exponential model. Hence, we used the linear model for non-evolving drift rates and an exponential model for evolutionary drift rates. To account for both kinds of drift rate behaviour, we used the following two models:

\begin{enumerate}
    \item \textit{Linear model of drift rates} \citep{2017ApJ...836..224M} \\
    
    This model assumes a linearly evolving drift rate ($D$) with respect to increasing pulse number. Thus, devising an equation which depends linearly on pulse number ($p$) since the onset of the drift sequence, we get 
    \begin{equation}\label{eqn:eqn1}
        D = \frac{d\phi}{dp} = a_1 p + a_2
    \end{equation}
    where $a_1$ and $a_2$ are constants, and $\phi$ is the pulse phase. We can integrate eqn. \ref{eqn:eqn1} to get the dependence of pulse phase on pulse number which would be a quadratic relationship. 
    \begin{equation}\label{eqn:eqn2}
        \phi(p) = a_1 p^2 + a_2 p + C
    \end{equation}
    
    Here $C = P_2 d + \phi_0$ is the integration constant that can be associated with a physical parameter, $P_2$, which is the ``horizontal'' separation between driftbands and is assumed to be a constant for the model fit of each drift sequence. Thus, the model has four free parameters, $a_1$, $a_2$, $a_3$, and $P_2$, to be accounted for.\\
    
    \item \textit{Exponential model of drift rates} \citep{2022ApJ...933..210M} \\
    
    For the cases where individual driftbands can show significant curvature, the exponential model can be a better fit. In this model, the driftbands are modelled with an exponential function which assumes an exponential decay rate for the drift rate, $D$
    \begin{equation}\label{eqn:eqn3}
        D = \frac{d\phi}{dp} = D_0 e^{-p/\tau_{\rm r}} + D_{\rm f}
    \end{equation}
    where $D_{\rm f}$ is the asymptotic drift rate, $D_0$ is the difference between $D_{\rm f}$ and the drift rate at the beginning of the drift sequence, $p$ is the number of pulses since the onset of the drift sequence, and $\tau_{\rm r}$ is the drift rate relaxation time (in units of the rotation period). To get the relationship with $\phi$ and $p$, we integrate eqn. \ref{eqn:eqn3}
    \begin{equation}\label{eqn:eqn4}
        \phi = \tau_{\rm r} D_0 \left( 1 - e^{-p/\tau_{\rm r}} \right) + D_{\rm f} p + \left( \phi_0 + P_2 d \right)
    \end{equation}
    where $\phi$ is the phase of a sub-pulse; $\phi_0$ is an initial reference phase; $d$ is the (integer) driftband number; and $P_2$ is the longitudinal spacing between successive driftbands. Thus, the model has five free parameters, $D_0$, $D_{\rm f}$, $\tau_{\rm r}$, $\phi_0$, and $P_2$, of which the expression $\phi_0+P_2d$ defines the pulse phase at $p = 0$. The last term in eqn. \ref{eqn:eqn4} is the same as the constant in eqn. \ref{eqn:eqn2}.
\end{enumerate}

Using eqn. \ref{eqn:eqn2} and \ref{eqn:eqn4} from the linear and exponential drift rate models, we employed either of the two on different mode sequences. The fitting procedure for either of the models was carried out following the method described in \cite{2022ApJ...933..210M}. 

\begin{figure*}
    \centering
    \includegraphics[width=\textwidth]{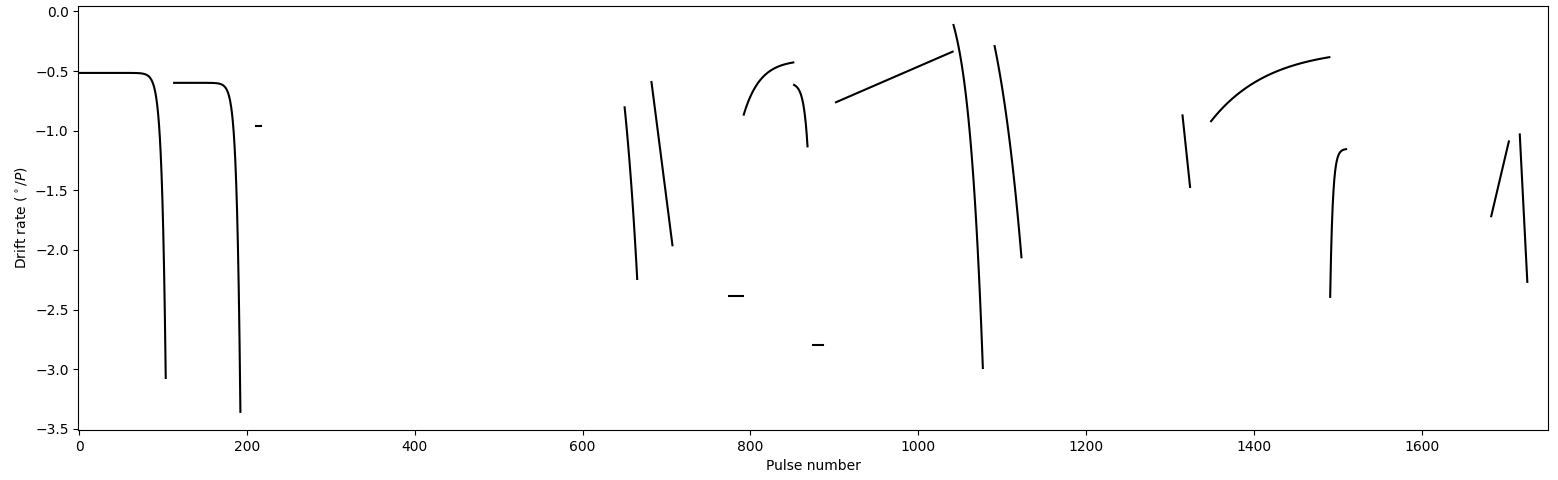}
    \caption{For the MJD 59529 scan 2 observation of J0026--1955, the above plot shows the drift rates as obtained using the linear and exponential drift rate models discussed in section \ref{bound}. Some of the corresponding driftband fits are shown in Fig. \ref{fig:figure2}.}
    \label{fig:drfit}
\end{figure*}

The pulsar exhibits both stable subpulse drifting (no evolution) and evolutionary drifting. Therefore, firstly we determined the mode boundaries (the beginning and end of a drift sequence) by visually inspecting the drift rate evolution in the pulse stack. Then, the evolutionary and non-evolutionary drift sequences were separated, with the caution that not too many mode boundaries are made. Sub-pulses were first smoothed using a Gaussian kernel of width $\sim3.6$ ms (i.e. $1^{\circ}$ of pulsar rotation, the approximate width of a subpulse). Then, the subpulses in each drifting sequence were identified by determining the peaks above a certain flux density threshold. This threshold was chosen such that for a minimum pixel value, no sub-pulse is identified in the off-pulse region. Each sub-pulse in the drift sequence is then assigned a driftband number. Finally, depending upon the drift rate model of choice, the driftbands are fitted (using SciPy’s \texttt{curve\_fit} method) with the functional form of the sub-pulse phases as mentioned in eqn. \ref{eqn:eqn2} and \ref{eqn:eqn4}. Examples of drift rate fitting are shown in Fig. \ref{fig:figure2}, where the bright diagonally arranged patterns are the driftbands and the white overlayed lines are the driftband fits. The drift rate evolution across an entire observation (scan 2 of observation made on MJD 59529) using the above methodology is shown in Fig. \ref{fig:drfit}. Here the $x-$axis shows the pulse number and the $y-$axis shows the drift rate in $^{\circ}/P_1$ units. The different curves correspond to either a linear or an exponentially varying drift rate across a drift sequence.

The models described above do not take into account the geometric curvature that must be present (to some degree), as discussed earlier. However, we argue that the geometric curvature must be negligible in J0026--1955's pulse window. As seen in Fig. \ref{fig:figure1} (and Fig. \ref{fig:figure2}), the pulsar exhibits a variety of drifting modes with inconsistent drift rates. If the geometric curvature was significant across the pulse window, it should be visible in all the modes, despite their evolutionary and non-evolutionary features. The fact that the characteristic `S'-shape of geometric curvature is not visible throughout leads us to conclude that it must be negligible across the pulse window for this pulsar. We, therefore, do not attempt to include geometric curvature in our models. In the next subsection, we describe the drifting modes and their various sub-classes, among which is a non-evolutionary mode (A0), in which the driftbands appear straight (see the left panel of Fig. \ref{fig:figure2}). This mode demonstrates the lack of a significant presence of geometric curvature.

\subsection{Drift Mode Classification}\label{modeclass}

\cite{2022ApJ...933..210M} categorised the drifting behaviour into two different classes: A and B. They further made sub-categories of modes A and B depending on the qualitative properties of drift sequences, their appearance and context. In this work, the broad classification into modes A and B follows \cite{2022ApJ...933..210M}, but our subcategories of these modes are completely different, and are based on the drift rate modulation instead of organisation in drifting patterns. In our analysis, we first modelled the drift rates exhibited by the pulsar and then categorised them. The drift rate modelling provided insight into the drifting behaviour, which was the basis for our mode classification, as described below.

\begin{figure}
    \centering
    \includegraphics[width=\columnwidth]{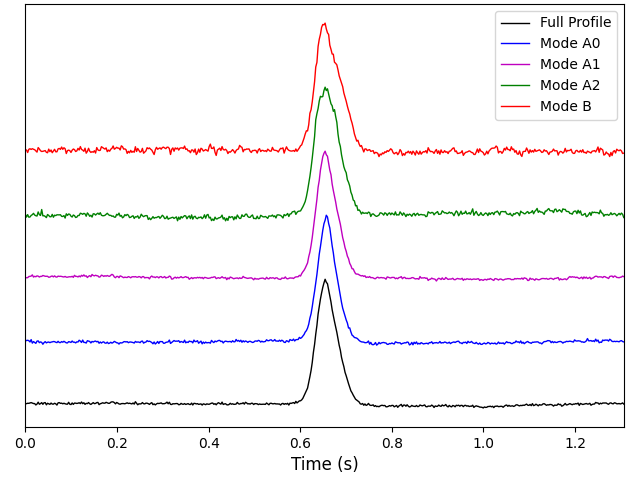}
    \caption{Average modal profiles of subpulse drifting modes exhibited by J0026--1955. Note that the average profiles of modes A0 and A1 are very similar. However, for modes A2 and B, the profiles are comparatively wider.}
    \label{fig:figure3}
\end{figure}

\begin{enumerate}
    \item \textbf{Mode A:} Mode A is classified as an umbrella mode category which encompasses the slower drift rates. The pulsar in mode A exhibits organised as well as unorganised driftbands. All of the evolutionary subpulse drifting behaviour exhibited by the pulsar can also be sub-categorised under mode A.

    \begin{enumerate}
        \item \textit{Mode A0:} This is the non-evolutionary sub-category of mode A. These are mode sequences which possess an almost constant drift rate and do not exhibit any evolutionary behaviour, as shown in Fig. \ref{fig:figure2}. Along with a stable drifting rate ($\sim -0.6^{\circ}/P_1$), mode A0 also has the largest mode length. The sequences, at times, do show frequent interruptions and rapid but temporary deviations in the drift rate. However, the overall drift rate still hovers around a constant number. The occurrence fraction of mode A0 in the complete set of observations was about 12\%.

        \item \textit{Mode A1:} According to our drift rate classification, sequences which demonstrate a slow evolution from fast to slow drift rates, as shown in Fig. \ref{fig:figure2}, are labelled as mode A1. In an extreme case of drift rate evolution in mode A1, the sequence begins with a small $P_3$ value of about 16$P_1$ and ends after 110 pulses with a much different $P_3$ of about 60$P_1$. Mode A1 also had the largest occurrence fraction of $\sim$17\% among all the subpulse drifting modes.

        \item \textit{Mode A2:} In addition to the evolution from faster to slower drift rates, the pulsar also exhibits the opposite evolutionary behaviour. Mode A2 sequences begin with a slow drift rate, where the driftbands are far apart and evolve towards a faster drift rate. In our data, mode A2 had a total occurrence fraction of $\sim$7\%. The sequences in mode A2 are generally short-lived and consist of 3-4 driftbands before the sequence ends with a faster drift rate (see fig. \ref{fig:figure2}). We also note that most occurrences of mode A2 are followed by a null. This possible correlation is discussed in detail in section \ref{sec:drnullcorr}. \cite{2022ApJ...933..210M} assumed this mode as a combination of mode A and the faster drifting mode (mode B). However, we believe that this is yet another evolutionary mode of J0026--1955, as the driftbands are fully connected throughout the drift sequence and exhibit a slow evolution rather than a sudden change in drift rate.
    \end{enumerate}

    As expected, from Fig. \ref{fig:figure3} it can be seen that the modal profiles of all the sub-categories of mode A show similar features despite the dissimilar drift rate behaviour, lending credibility to our classification scheme. \\

    \item \textbf{Mode B:} The pulsar also exhibits a faster drift rate on its own without being a part of any evolutionary behaviour. This was present in only $\sim$4\% of the total observation. Mode B sequences are short-lived, are generally isolated occurrences, and do not show a drift rate evolution, as shown in Fig. \ref{fig:figure2}. Mode B sequences can be found anywhere in the pulse stack, even in the midst of long, otherwise uninterrupted null sequences. The average drift rate for mode B is $\sim -1.6^{\circ}/P_1$. The average modal profile of mode B is shown in Fig. \ref{fig:figure3}, which shows slightly different features with a skewed average profile, as compared to mode A profiles.
\end{enumerate}

An additional feature was sometimes noted in the drift sequences of J0026--1955, where an extra driftband seems to appear towards the leading edge. An example can be seen at around pulse number 2125 in panel (c) of Fig. \ref{fig:figure2}, in mode A2. There is a sudden break in the middle of the driftband, and both pieces look disassociated. It appears that towards the end of the first driftband, the drift rate fastens, and for the second driftband, the drift rate begins at a faster rate and then slows down. Overall, if the sudden drift rate change and the break are ignored, they seem to form a full driftband. During our analysis, we have not accounted for the break and considered the driftband in full wherever such a deviation was noted.

\begin{figure}
    \centering
    \includegraphics[width=\columnwidth]{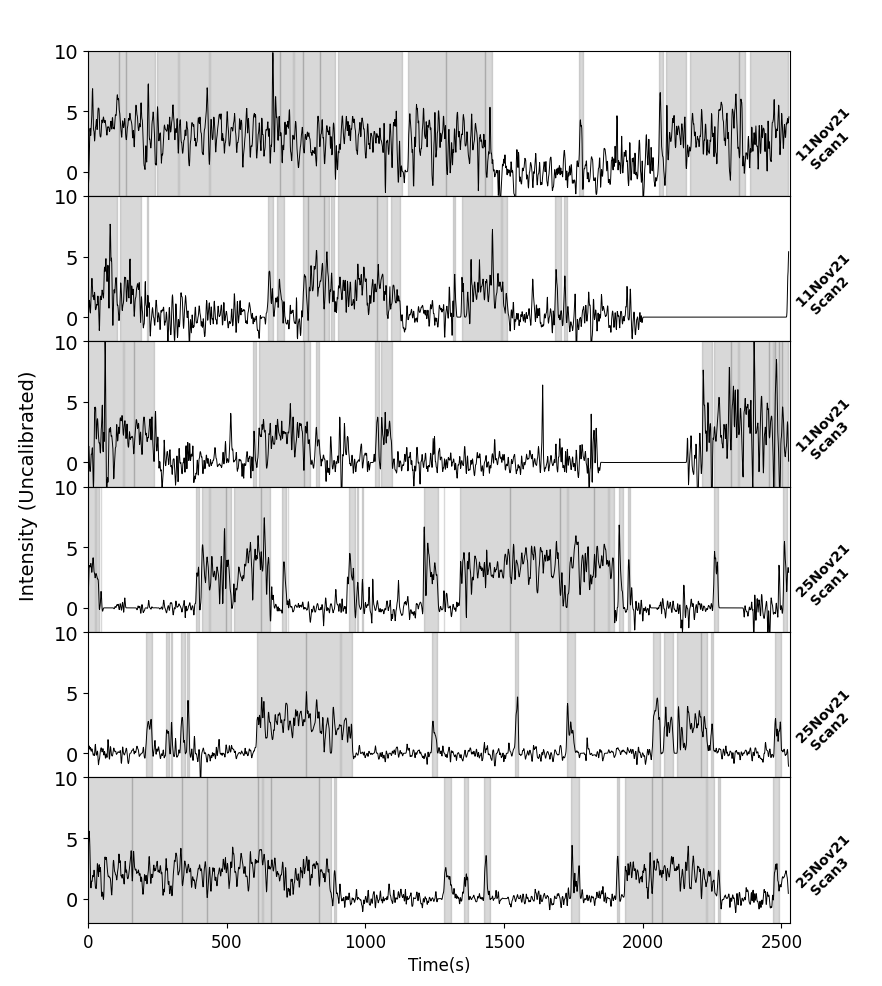}
    \caption{Peak flux density within the on-pulse window as a function of time for  the six uGMRT observations at 400 MHz. The shaded grey regions show detected subpulses. The unshaded region corresponds to nulls and RFI occurrences. The peak fluxes were measured after smoothing the time series with a Gaussian filter (1$\sigma$) to suppress the noise contribution and accentuate the contrast between nulling and burst sequences.}
    \label{fig:figure4}
\end{figure}

\subsection{Nulling}\label{null}

Nulling is the temporary disappearance of emission from a pulsar for brief periods of time. After deciding the mode boundaries using the method described in \ref{modeclass}, sequences with no subpulse detection were counted as nulls. In our observations, PSR J0026--1955 showed evidence of both long bursts of pulses and long nulls. The long nulls are sometimes interrupted with short mode B sequences. Fig. \ref{fig:figure4} shows the entire length of null and burst sequences in all our observation scans. The longest null sequence goes on for 1117 pulses ($\sim$25 minutes). In contrast, the most prolonged burst in our observations lasts for 867 pulses ($\sim$19 minutes). 

Fig. \ref{fig:figure4} shows intensity as a function of time for all the observations. The shaded grey regions show detected subpulses, and the rest are nulls. The degree of nulling in a pulsar can be quantified in terms of the nulling fraction (NF), which is the fraction of pulses with no detectable emission. Overall, J0026--1955 was in a null state for more than half of our observations, with an estimated total nulling fraction of $\sim$58\%. This differs from the nulling fraction of $\sim$77\% obtained at 155 MHz using the MWA \citep{2022ApJ...933..210M}. This discrepancy and the nulling behaviour of J0026--1955 are discussed in detail in section \ref{disc:null}.

\subsection{Drift Rate Evolution}\label{drevol}

PSR J0026--1955 exhibits multiple drift rates and evolutionary features throughout the drift sequences and individual driftbands. Initially, we used the linear and exponential models to understand the drift rate behaviour within a sequence, as described in section \ref{bound}. However, driftbands and sequences in J0026--1955 exhibit more complicated evolutionary features, as seen in the mode A1 and A2 occurrences in Fig. \ref{fig:figure2}, where even an exponential model fails to accurately describe the evolutionary drift rates correctly. 

\begin{figure*}
    \centering
    \includegraphics[width=0.33\textwidth]{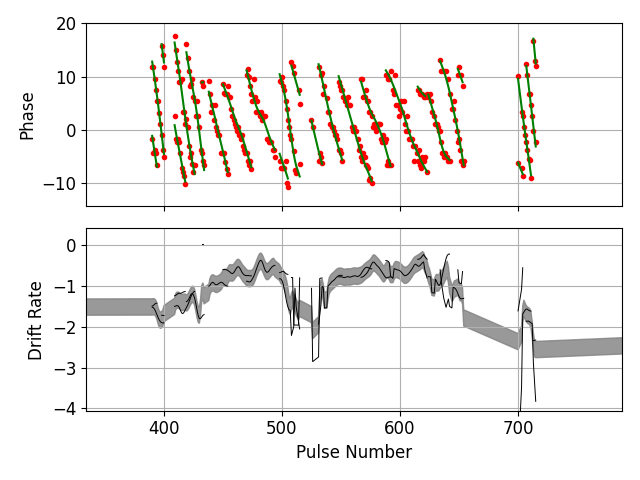}\includegraphics[width=0.33\textwidth]{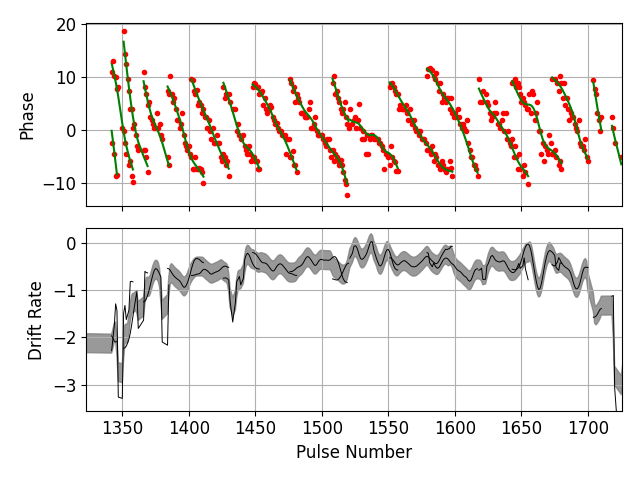}\includegraphics[width=0.33\textwidth]{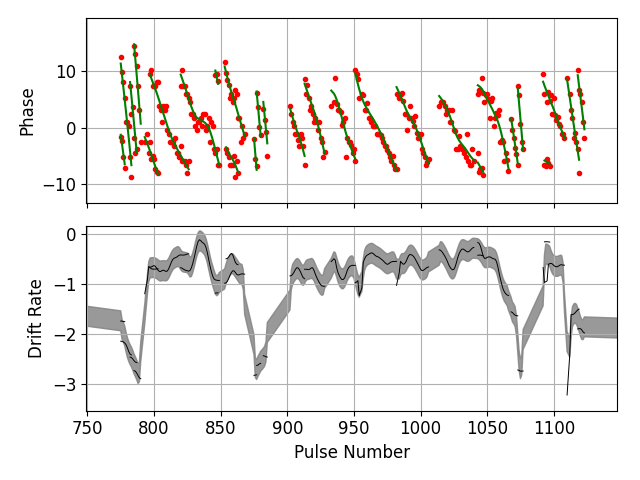}
    \caption{In the top panels of all sub-figures, the red dots show the location of subpulses, while the green lines correspond to fits to driftbands using a cubic smoothing spline method. In the lower panels, the black solid lines show the drift rates ($^{\circ}/P_1$) of individual driftbands, while the solid grey line shows the average drift rate. The figures represent only a few drift sequences.}
    \label{fig:figure6}
\end{figure*}

We further explored the drift rate behaviour of J0026--1955 by studying the variation in drift rate with each pulse in a driftband. To calculate the evolution of the drift rate with each pulse, we followed the methodology described in \cite{2022ApJ...934...23S}. We first employed a cubic smoothing spline estimate using the \textit{SmoothingSplines}\footnote{\url{https://github.com/nignatiadis/SmoothingSplines.jl}} software package on each of the driftbands, irrespective of their mode identity or drift mode boundaries. Then, we obtained the drift rate (phase/pulse number) by calculating the gradient of the fitted spline function at every pulse number for each driftband. As seen in J0026--1955 pulse stacks, there can be two driftbands at a given pulse number. Fig. \ref{fig:figure6} shows examples of some drift sequences, where the top panel shows part of the pulse stack; where the red dots indicate the location of subpulses. Here, the green line is the cubic spline fit, which was obtained with the smoothing parameter $\lambda = 100$. In the bottom panel of Fig. \ref{fig:figure6}, the black lines show the gradient calculated at each pulse number for every driftband. As some pulses contain two subpulses, yielding two measurements of the drift rate for that pulse number, we estimate multiple drift rates at some of the pulse numbers. The solid grey envelope shows the mean drift rate at every pulse number where the contribution from multiple driftbands at any given pulse is averaged. 

\begin{figure}
    \centering
    \includegraphics[width=\columnwidth]{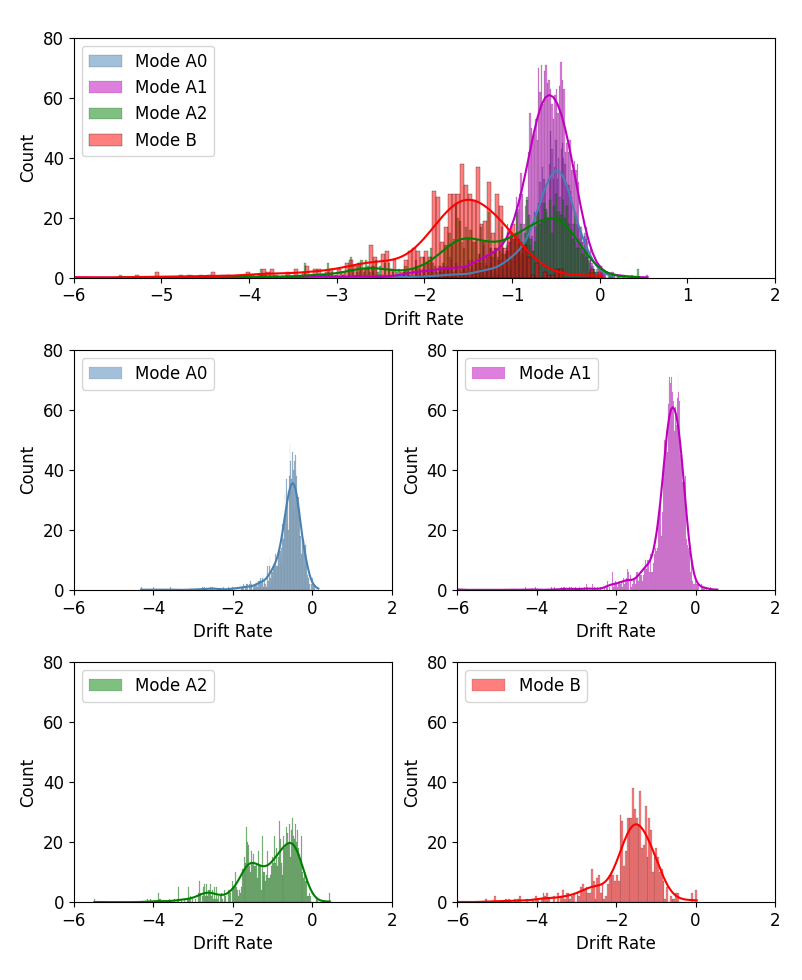}
    \caption{The figure shows a distribution of drift rates ($^{\circ}/P_1$) for different modes. The topmost panel shows the combined distribution and the lower panels show the drift rates for individual modes.}
    \label{fig:figure7}
\end{figure}

A subset of pulsars that exhibit multiple subpulse drifting modes shows a harmonic relationship between $P_3$, as reported in previous studies \citep{2013MNRAS.433..445R,2019ApJ...883...28M}. In the case of J0026--1955, a similar analysis with drift rates leads to interesting implications. A cumulative and modal histogram of individual drift rates obtained by taking a gradient of the smoothing spline fit of driftbands at each pulse number can be seen in Fig. \ref{fig:figure7}. The top panel shows a distribution of all drift rates, where each colour corresponds to the different modes classified for J0026--1955. The lower panels of Fig. \ref{fig:figure7} display the distribution of drift rates for all observed subpulse drifting modes of J0026--1955. Apart from two distinct drift rate peaks at approximately $-$0.5$^{\circ}/P_1$ and $-$1.6$^{\circ}/P_1$, a third peak is visible at around $-$2.7$^{\circ}/P_1$. The peak values of the drift rates form an arithmetic sequence with a common difference of approximately $-$1.1$^{\circ}/P_1$. Further, considering eqn. 1 and 2 in \cite{2019ApJ...883...28M}, it can be implied that if an arithmetic spacing exists between drift rates, then the corresponding number of sparks will also have an arithmetic relationship, assuming a constant carousel rotation rate ($P_4$).

\subsubsection{A fourth-order polynomial fit of drift rates}

\begin{figure*}
    \centering
    \includegraphics[width=\textwidth]{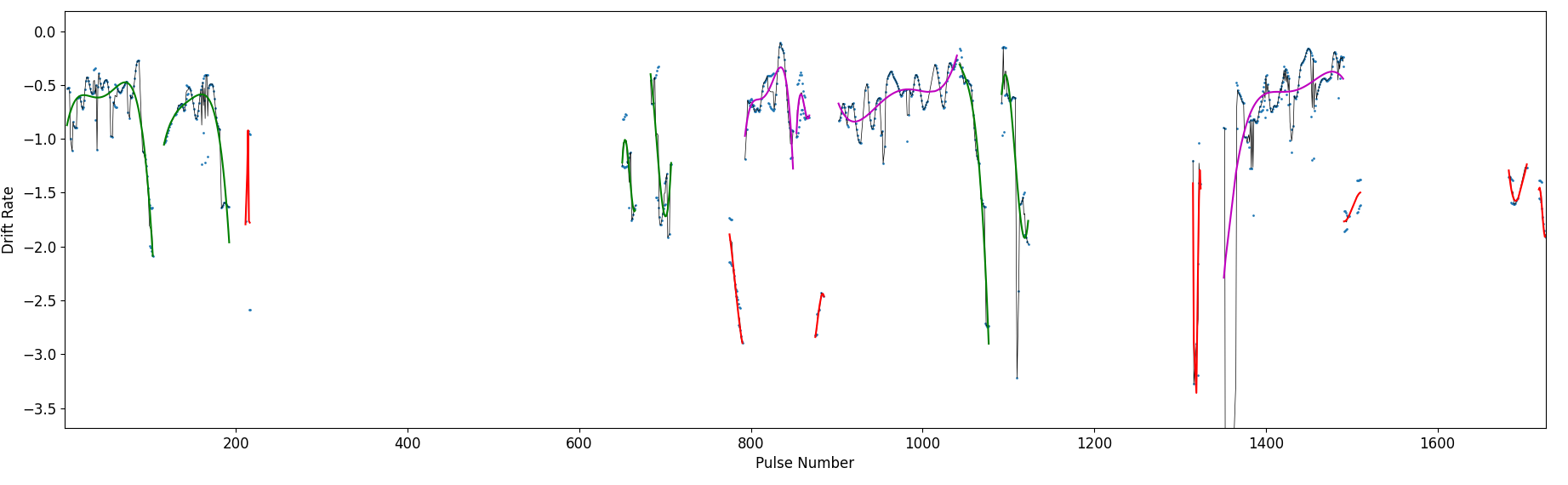}
    \caption{For the same observation of J0026--1955 shown in Fig. \ref{fig:drfit}, the above plots show the fourth-order polynomial drift rate model fits. Using the cubic smoothing spline method, we calculated the drift rate ($^{\circ}/P_1$) at each pulse number which is shown using the blue dots. The black line shows the average drift rate per pulse number. Drift rates for each of the modes are fit using a fourth-order polynomial, which describes a global evolution of drift rates. The colour scheme is similar to Fig. \ref{fig:figure7}}
    \label{fig:figure8}
\end{figure*}

On a closer examination of Fig. \ref{fig:figure6}, it is evident that the drift rate is irregular. Furthermore, the drift rate does not simply evolve towards a higher or lower rate but shows variability, even within a particular mode. A linear or exponential drift rate could not accurately comprehend the complexity of this drift rate evolution. Hence, we tried to fit the average drift rate with a polynomial. Employing the polynomial regression method using \texttt{scikit\_learn}, we successfully fit a fourth-order (quartic) polynomial function to the evolving drift rates. A higher-order polynomial could also describe the evolutionary drift rate. However, such a model might be counter-intuitive and would only provide a customised fitting rather than a general model. Fig. \ref{fig:figure8} shows the fourth-order polynomial fit of the drift rates (blue dots) for scan 2 of the observation made on MJD 59529 (November 11, 2022). The black line is the average drift rate at each pulse number, same as the grey envelope in Fig. \ref{fig:figure6}. Different colours of the fourth-order polynomial fit correspond to different modes, following the colour scheme of Fig. \ref{fig:figure7}. A direct comparison between the drift rate models in Fig. \ref{fig:drfit} and Fig. \ref{fig:figure8} can be made, where the latter shows the evolution of drift rate within a drift mode sequence. The fourth-order polynomial model can be seen to better describe the drift rate modulation as compared to the more simplistic, linear and exponential models, which lacked the detail.

\begin{figure*}
    \centering
    \includegraphics[width=.33\textwidth]{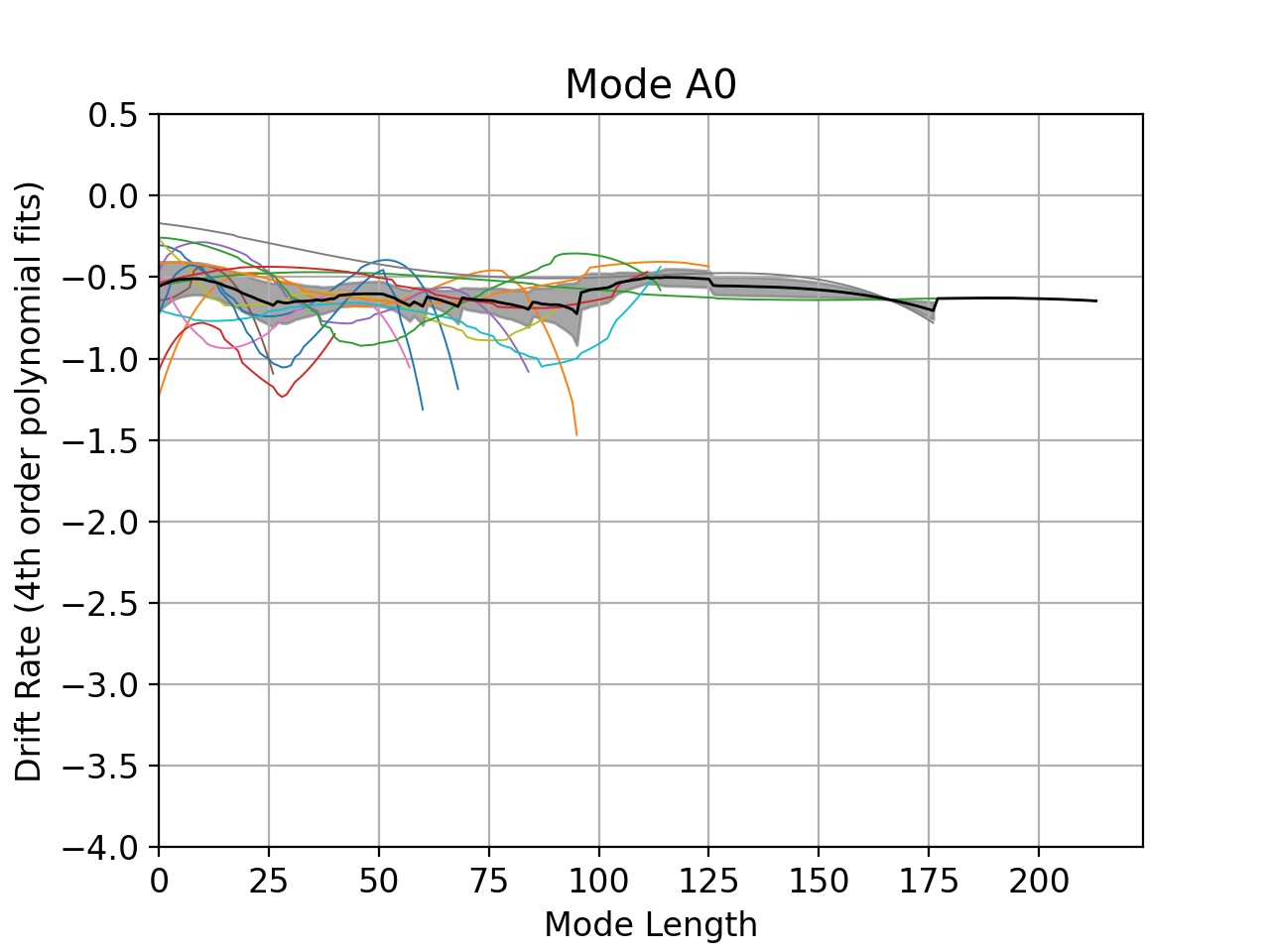}\hfill
    \includegraphics[width=.33\textwidth]{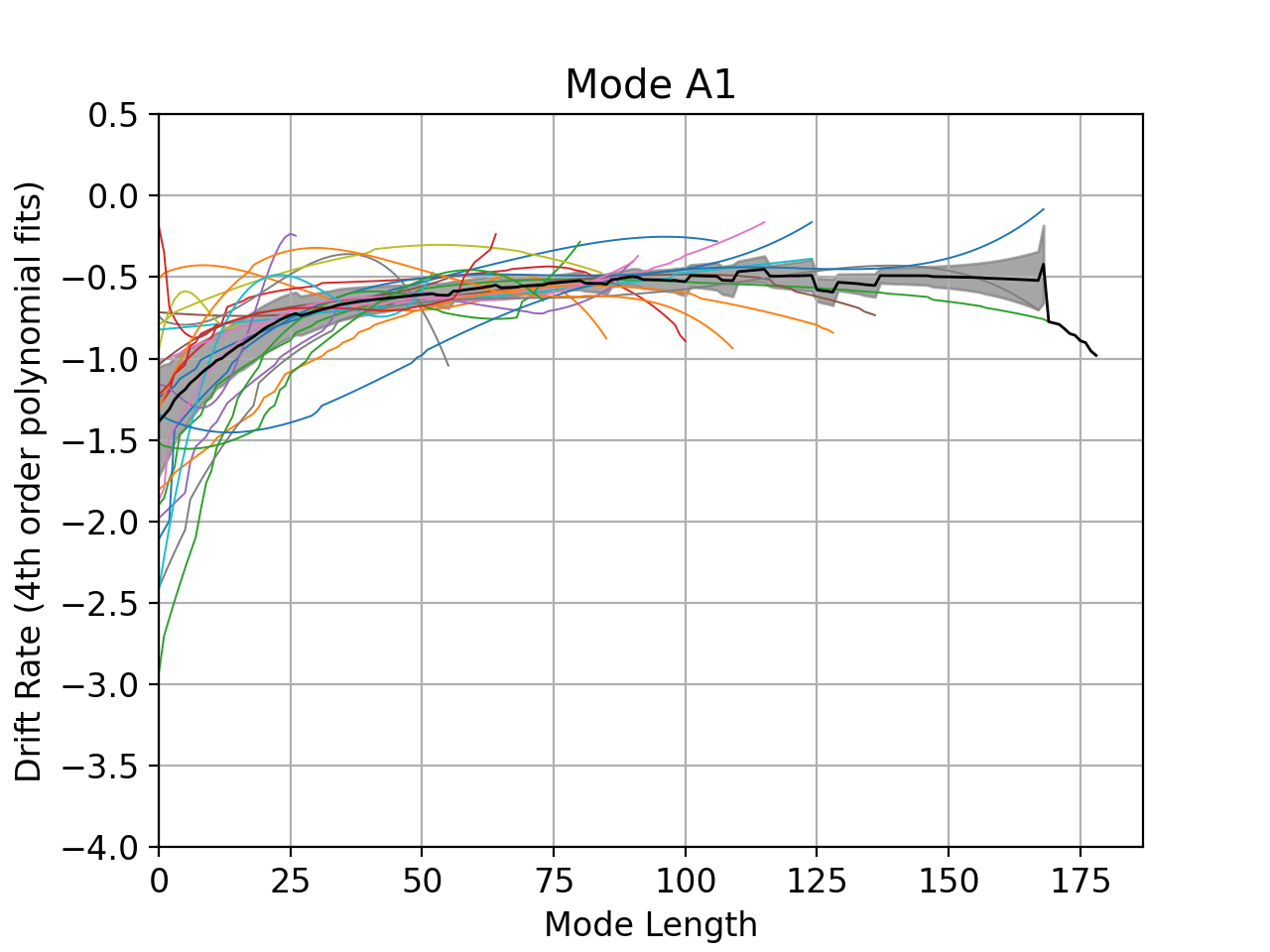}\hfill
    \includegraphics[width=.33\textwidth]{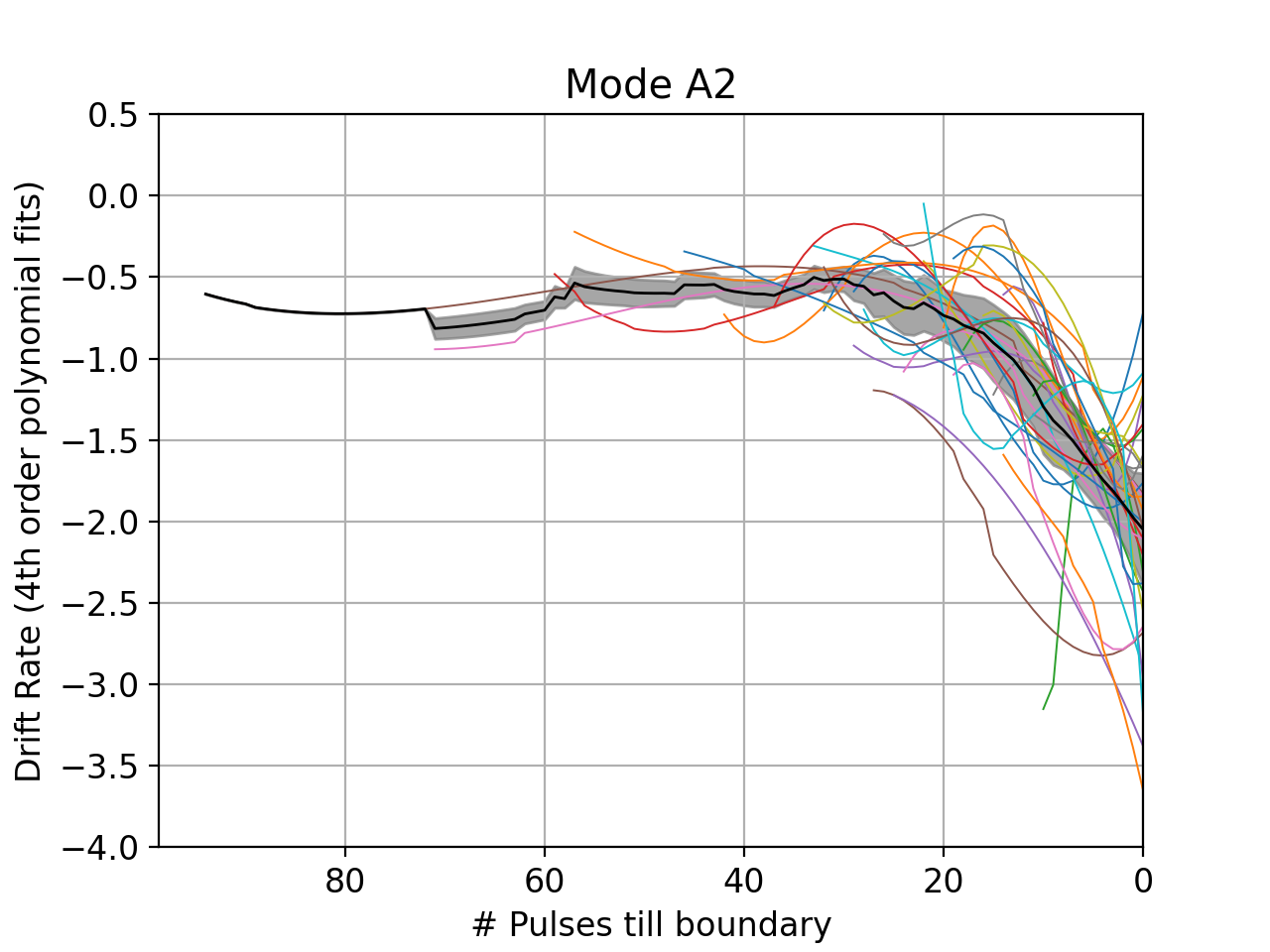}
    \caption{The figure combines all the drift rate fits using the fourth-order polynomial. The black line shows the average of all drift rates, and the grey envelope shows the $1\sigma$ deviation around the mean. \textit{Left:} Drift rate for mode A0 stays constant on average despite sudden changes, as shown by the black line. \textit{Middle:} Mode A1 is an evolutionary mode where the drift rate evolves from faster to slower drift rates. This is also shown by the black line, which evolves from an average drift rate of $\sim$$-$1.3$^{\circ}/P_1$ to $\sim$$-$0.5$^{\circ}/P_1$ with pulse number. \textit{Right:} The evolutionary behaviour of mode A2 is the opposite of mode A1, where it evolves from a slower to a faster drift rate towards the end. In this plot, we have plotted the drift rates such that all drift sequences end together. This exercise shows the claimed drift rate behaviour. Following the trend of the black line (average drift rate), one can note that the drift rate evolves from $\sim$ $-$0.5$^{\circ}/P_1$ to $\sim$ $-$2.0$^{\circ}/P_1$.}
    \label{fig:figure9}
\end{figure*}

To understand the overall drift rate modulation in various modes, we overlayed the drift rate fits for each mode, as shown in Fig. \ref{fig:figure9}, where the $x-$axis shows mode length and $y-$axis the drift rate. The black line in each subplot shows the mean of drift rates (from the polynomial fits) with the pulse number. The grey envelope corresponds to the $1\sigma$ deviation from the mean drift rate. As expected, mode A0, which does not show any noticeable evolutionary behaviour had an almost constant average drift rate across all instances, around -0.6. In contrast, the drift rate evolution of modes A1 and A2 is observable in their respective subplots. Drift rates in mode A1 can be seen to evolve towards a slower drift rate as compared to the commencing rate, whereas mode A2 shows an overall evolution towards a faster drift rate as it reaches the end of a drift sequence.

\subsection{Memory across nulls}\label{sec:man}

\begin{figure}
    \centering
    \includegraphics[width=\columnwidth]{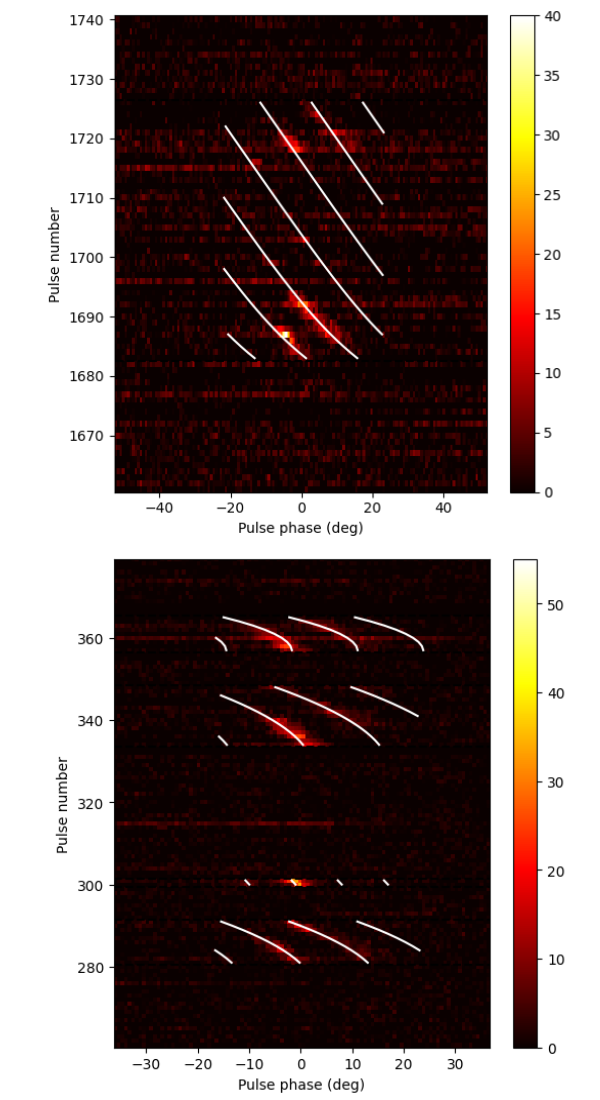}
    \caption{Examples showing speculative drifting during nulls for J0026--1955. \textit{Top:} Between the two drifting sequences, there is a single burst at 1703. Extrapolating the drift rates of the first sequence (pulse 1683 to 1694) aligns well with the drifting behaviour at later pulse numbers. The driftband fits are shown to illustrate the approximate locations of the missing driftbands. \textit{Bottom:} Between the four drifting sequences, though the drift rates may not align, the phase memory of the last burst before nulls is almost retained.}
    \label{fig:figure5}
\end{figure}

We also investigated the presence of memory across nulls, as reported by \cite{2022ApJ...933..210M}. Our analysis shows that the short-lived nulls of PSR J0026--1955 indicate evidence of subpulse memory across nulls. This could be a subpulse phase memory or a drift rate memory. In the first scenario, the phase of the last subpulse before the null and the first subpulse after the null are similar. On the other hand, a drift rate memory could be a case where the drift rate across the null stays the same, and the driftbands could be extrapolated. 

We followed the model fitting technique described in \ref{bound} to explore the latter. To check if there is indeed a drift rate connection between the sequence before the null and the sequence after the null, we fit the previous drift sequence (before null) using the model of choice (following the method in \ref{bound}). We then extrapolate the model to a subpulse right after the null sequence ends, allocating the subpulse a reasonable driftband number. We can consider a drift rate memory if the projected phase of the subpulse after the null matches the real subpulse within an error range. The error on phase prediction is calculated from the covariance matrix of the model fit (using standard uncertainty propagation). If the phase of the real pulse falls outside the subpulse phase range projected by the model, then we consider that there is no memory across the null. On the other hand, if the projected phase and the phase of the real pulse are within the error bars and smaller than $P_2$ (phase distance between two subpulses within a pulse), then we classify that null as being consistent with being phase-connected subpulse across the null. An example of drift rate memory across nulls is shown in the top panel of Fig. \ref{fig:figure5}, where the white lines depict the drift rate fits. The top panel in Fig. \ref{fig:figure5} shows two drift sequences on either end and a null. By fitting the sequence before the null and projecting the drift rate behaviour to the latter sequence, one can note that the drift sequences before and after the null are consistent. 

We, however, do not find many instances of subpulse phase memory across nulls in our data. A handful of instances were initially recognised by visual inspection, as they did not show a drift rate memory across nulls. They were further investigated by calculating the phase of the subpulse before and after null. One such example is shown in the bottom panel of Fig. \ref{fig:figure5}, where the subpulse phase before and after the null are almost the same. In contrast, since the sequences have different drift rates, a drift rate memory might not be present. A more careful analysis of longer observations is required to verify this fully.
\section{Discussion} \label{sec:discussion}

PSR J0026--1955 exhibits a multitude of pulse-to-pulse modulation phenomena, viz. subpulse drifting, mode switching, and nulling, as shown in Fig. \ref{fig:figure1}. For each of the driftbands, subpulses arrive at earlier phases with pulsar rotation, thus conforming to the `positive drifting' class of subpulse drifters \citep{10.1093/mnras/sty2846}. However, the vertical separation between driftbands (i.e., $P_3$) can be seen to vary between different sequences, as well as within a drift sequence. There are instances where J0026--1955 can be seen to abruptly switch from one subpulse drifting mode to another, a behaviour exhibited by many other pulsars \citep[e.g.,][]{2019ApJ...883...28M, 2022MNRAS.509.4573J}. However, in addition to the abrupt mode change, the drift rate can also sometimes be seen to evolve gradually, which is a relatively rare phenomenon, observed in only a small subset of subpulse drifting pulsars, e.g., PSR B0943+10 \citep{ 2018A&A...616A.119B}, PSR B0809+74 and PSR B0818--13 \citep{10.1093/mnras/204.2.519}. For PSR J0026--1955, a quick glance at mode A1 and A2 sequences (evolutionary drift modes) in Fig. \ref{fig:figure2} shows this variable drift rate. The pulsar also exhibits long and short-duration nulls, where an association between drifting and nulling can be drawn. Below, we discuss the variety of phenomena exhibited by J0026--1955 in light of the analysis presented in section \ref{sec:analysis}.

\subsection{Nulling Behaviour of J0026--1955} \label{disc:null}

The analysis presented in section \ref{null} highlights the unique nulling behaviour of J0026--1955, which exhibits complex emission properties. The nulling fraction at 400 MHz is approximately 58\%, an estimate reached from 330 minutes of observation, whereas, at 155 MHz, it was estimated to be 77\% from 192 minutes of MWA observation \citep{2022ApJ...933..210M}. This discrepancy suggests a possibility that the nulling behaviour of J0026--1955 may not be broadband and that there may be frequency-dependent mechanisms at play. Frequency-dependent nulling has been observed in some pulsars, where the nulling behaviour varies depending on the observed frequency \citep[e.g.,][]{2007A&A...462..257B}. However, given the modest separation in the frequency bands (with band edges separated by $\sim$130 MHz and the centre frequencies differing by a factor of $\sim$2.5), it is unclear if the observed discrepancy is entirely attributable to frequency-dependent nulling. Alternatively, the observed inconsistency could simply be an unintended observational bias, where the MWA observations were incidentally made around the long nulls. The perceived discrepancy could also be because of the presence of an emission component with a shallow spectral index leading to a null at 155 MHz. Assuming that J0026--1955 exhibits broadband nulling, a combination of the number of nulls out of the total number of pulses observed at 155 MHz and 400 MHz will imply a nulling fraction of $\sim$65\%.

It is also possible that the nulling behaviour of J0026--1955 is complex and multi-faceted, involving both broadband and frequency-dependent mechanisms. Different models attribute pulsar nulling to intrinsic changes in the magnetosphere, such as temperature fluctuations altering coherence conditions \citep{1981IAUS...95...99C,1986ApJ...300..540D}, switching between gap discharge mechanisms \citep[e.g.,][]{1986ApJ...309..362D,1997ApJ...491..891Z}, variations in magnetospheric currents \citep{2010MNRAS.408L..41T}, change in pulsar beam geometry \citep{2007MNRAS.380..430H,2009MNRAS.393.1391H,2008MNRAS.385.1923R}, disruption of the entire particle flow in the magnetosphere \citep{2006Sci...312..549K}, etc. Though such models may be able to describe the long-period nulls seen for J0026--1955, the presence of subpulse (phase or drift rate) memory across nulls challenges these theories for at least the case of short nulls where such memory exists. Further investigation is needed to fully understand the nature of the observed disparity and the complex emission properties of J0026--1955. 


\subsection{Drift Rate - Nulling Correlation}\label{sec:drnullcorr}

There have been only a limited number of investigations that explored a correlation between nulling and subpulse drifting. For example, PSR B0818--41 presents a case where a decrease in the pulsar drift rate is accompanied by a gradual decrease in intensity before the onset of a null \citep{2010MNRAS.408..407B}. PSR B0809+74 also shows an association between nulls and subpulse drifting, where the drift rate deviates from normal after the nulls \citep{2003A&A...399..223V}. Using the partially screened gap model, the authors suggest that some kind of ``reset'' of the pulsar’s radio emission engine occurs during the nulls, which is responsible for the conditions of the magnetosphere. For PSR B0809+74, \cite{2003A&A...399..223V} suggest that nulling and subpulse drifting may be related, with changes in emission beam geometry potentially causing both the nulling and changes in the drift behaviour. \cite{2007MNRAS.377.1383W} suggested that emission can cease or commence suddenly when the charge or magnetic configuration in the magnetosphere reaches the so-called ``tipping point''; however, the triggering mechanism for such stimulus is unknown. In the case of PSR J1822--2256, \cite{2022MNRAS.509.4573J} also showcase a relationship between nulling and mode changing, where a null preceded most occurrences of their mode D. Such correlations suggest that emission mechanisms and magnetosphere dynamics between nulls and subpulse drifting may be strongly related.

PSR J0026--1955 also provides some compelling evidence of a possible correlation between subpulse drifting and nulling. Our observations suggest a complex and dynamic mechanism underlying the pulsar radio emission, with important implications for understanding astrophysical processes in extreme environments. In particular, we have found that the pulsar J0026--1955 likely switches to a null state after mode A2. Mode A2 is an evolutionary mode, where the drifting begins at a slower drift rate and evolves towards a faster drift rate before the mode eventually ends. In 27 out of 31 instances, mode A2 is followed by a null, either short or long. In such cases, the ramping up towards the faster drift rate begins remarkably consistently about 20 pulsar rotations prior to the null (see Fig. \ref{fig:figure5}), which suggests that the null itself might be causally related to the preceding drifting behaviour. In the rest of the four sequences, mode A2 was once at the end of our observation and was followed by either mode A0 or A1 in three instances. These occurrences can also be seen as a drift reset, where the pulsar temporarily evolves to a faster drift rate and then returns to a slower one. This reset could also be due to a change in the magnetospheric conditions before a ``tipping point'' was reached. It was also noted that not every null sequence followed mode A2, but a null followed most occurrences of mode A2. We did not find any strong correlation between the intensity of subpulses towards the onset or end of a null. The variation in pulse intensity of transitions from burst to null was sometimes abrupt and smooth at other times and lacked any compelling evidence of intensity dependence. It is worth noting that this kind of phenomenon, where nulls affect the drifting behaviour leading up to them, is relatively rare in contrast with the more commonly studied effect of nulls on burst sequences that come after them \citep[e.g., ][]{10.1093/mnras/204.2.519,2003A&A...399..223V,2022MNRAS.509.4573J}. A deeper understanding of these rare cases can provide useful insights into the complex dynamics relating nulls and subpulse drifting.


\subsection{Memory Across Nulls}

Only a limited number of pulsars retain the information about previous subpulses during nulls, as demonstrated by, e.g., \cite{1978MNRAS.182..711U} and \cite{2017ApJ...850..173G}. This memory retention can provide valuable insights into the true nature of the nulling phenomenon and any correlation with subpulse drifting it may have. If intrinsic changes in the pulsar magnetosphere lead to both nulls and drift-rate modulations, studying the interactions between these phenomena could offer valuable insights into the mechanisms that trigger and facilitate transitions between the different states. Subpulse memory across nulling in drifting pulsars has been generally attributed to either of the two reasons: (1) the polar cap continues to discharge, but the emission is not observed \citep{1982ApJ...263..828F}, or (2) the subpulse ceases drifting for the duration of the null \citep{1978MNRAS.182..711U}. In the first case, the absence of radio emission is attributed to the lack of coherent structure, and the drift rate remains the same before and after the null. Thus, a part of the drift sequence would be missing, though one could still map the driftbands with similar drift rates pre- and post-null, showing `drift rate' memory across nulls. Whereas in the second scenario, the subpulse drifting is thought to resume at the phase where it left off, demonstrating a subpulse `phase memory' across nulls. As described in section \ref{sec:man} have encountered possibilities of both `drift rate' memory and `subpulse phase' memory across nulls in the case of J0026--1955.

\subsubsection{Drift rate memory}

J0026--1955 exhibits a clear case of subpulse drift memory (top panel in Fig. \ref{fig:figure5}), where the pulsar seems to remember the drift rate even after a short ($\sim$20 pulses) null. The study described in \cite{1982ApJ...263..828F} suggests that nulling in pulsars results from an uninterrupted and stable discharge in the polar gap rather than a complete cessation of sparks. 
This observation is similar to the case of J1840--0840, where the drift rate stays the same across the null, and an entire driftband is missing from the sequence \citep{2017ApJ...850..173G}. PSR J1840--0840 also shows `subpulse phase' memory in conjunction with the `drift rate' memory. 

The cause of undetectable radiation during nulling can be attributed to the absence of the dominant coherence mechanism present during regular pulsar operation rather than the lack of particle flux from the polar cap \citep{1982ApJ...263..828F}. Therefore, during the null states, the subpulses on the polar cap may continue to drift during the null state, either at similar or different rotation speeds. In this case, the phase of the subpulse after the null sequence can be anticipated based on the duration of nulls. In the case of J0026--1955, when the pulsar switches its emission back on after a null, the phase of the subpulse can be extrapolated from the drift rate model before the null. The predicted phase was well within the error range for drift sequences around short nulls. This finding supports the idea that drifting and sparking may still be operational during nulls, providing further evidence for the two scenarios previously observed in drifting pulsars.

\subsubsection{Subpulse phase memory} 

As shown in the lower panel of Fig. \ref{fig:figure5}, J0026--1955 also presents evidence of a `phase' memory across nulls, where the drift rate does not necessarily stay the same across the nulls. Still, the phase of the last subpulse before the null is almost identical to that of the first subpulse after the null. Similar behaviour is observed in a few other pulsars like B0809+74 and B0818--13, where the drift rate changes after null sequences \citep{10.1093/mnras/204.2.519,2003A&A...399..223V,2004A&A...425..255J}. During some of these interactions, the pulsars also seem to indicate some kind of phase memory, such that information regarding the phase of the last subpulse was retained during the null state. \cite{1980ApJ...235..576C} propose that during the null, the drifting stops and the position of the sparks are remembered by the presence of a hotspot on the pulsar surface. This would imply that once the drifting resumes, the sparks will reform at their previous position. PSR B0031--07 was shown to retain the memory of its pre-null burst phase across short nulls \citep{1997ApJ...477..431V,2000MNRAS.316..716J}. In their study by \cite{2017ApJ...850..173G}, PSR J1840--0840 presents a unique example of both `subpulse phase' and `drift rate' memory across null. In our data, we only had a few occurrences of memory across nulls of both kinds. A more detailed analysis of drift rate and subpulse phase memory around nulls needs to be conducted for J0026--1955 using longer observations, thus increasing the sample size of such events.


\subsection{Subpulse Drifting Model for J0026--1955} \label{model}

J0026--1955 presents an exciting and rare phenomenon of drift rate evolution (modes A1 and A2), in addition to regular subpulse drifting with a constant drift rate (modes A0 and B). The pulsar exhibits changes in drift rate over sequences, with mode A1 showing a trend towards a slower drift rate and mode A2 showing a trend towards a faster drift rate with increasing pulse number. Furthermore, irrespective of the modal transitions, we have also observed an inter-modal driftband connectivity in most mode-switching cases. We utilised the techniques described in section \ref{drevol} to analyse the evolution of drift rates across the driftbands and sequences. Measuring the slopes of individual driftbands was essential to scrutinise any changes in the drifting pattern. 

Furthermore, we modelled the drift rate behaviour across the drift sequences using a quartic polynomial. The quartic model fits were then used to understand the global drift rate variation for the different modes, as shown in Fig. \ref{fig:figure9}. The figure shows that the modes A1 and A2 for J0026--1955 display a gradual evolution from their initial drift rates. It is noteworthy that the change of drift rate for mode A2 ($\sim$ $-0.5^{\circ}/P_1$ to $\sim$ $-2.0^{\circ}/P_1$) is almost twice as much as the change of drift rate in mode A1 ($\sim$ $-1.3^{\circ}/P_1$ to $\sim$ $-0.5^{\circ}/P_1$). Hereafter, we discuss the possible modifications/additions to the existing carousel model that can explain the observed unique behaviour of J0026--1955.

\subsubsection{Variable Spark Configuration in Carousel Model}\label{model1}

As discussed in \cite{10.1093/mnras/stt739}, \cite{2017ApJ...836..224M}, and \cite{2022MNRAS.509.4573J}, different modes and drift rates of a pulsar can be attributed to a carousel with varying numbers of sparks for each subpulse drifting mode. In such cases, the drift rate is observed to change abruptly. This sudden change cannot be ascribed to the carousel rotation rate, as it would imply significant magnetosphere reconfiguration over a short time scale. However, the drift rate change within a single pulsar rotation can be due to the reconfiguration of spark distribution. In the case of the evolutionary drifting modes of J0026--1955, the idea of a carousel with a constantly changing number of sparks to describe the changing drift rate may be counter-intuitive. Nevertheless, the non-evolutionary modes (A0 and B) may still have a fixed spark configuration.

An alternative hypothesis which could involve a changing carousel rotation rate, would need a slow change in the spark carousel itself, where the gradual evolution is a signature of the spark configuration relaxing into a new arrangement after a spark suddenly appears or disappears from the carousel \citep{2019PhDT.......160M}. Such ``relaxation'' of the drift rate is reminiscent of the behaviour observed in PSR B0809+74 \citep{10.1093/mnras/204.2.519}, where, after a null, the pulsar would temporarily attain a faster drift rate before relaxing into a steady drift rate. They further suggest that a perturbation might alter the drift rate (and emission, thus causing a null), after which the drift rate recovers exponentially to its normal value. 

If the number of sparks in the carousel is changing, then the carousel will take some finite amount of time for the sparks to rearrange themselves, for example, into the new configuration in which the sparks are equidistant from each other. During this relaxation time, the angular speed of individual sparks may differ from the angular speed of the whole carousel. In that case, the observed drift rate at any one moment will only depend on the spark that is ``under'' the line of sight during any given rotation period. In this view, the observed drift rate can appear to change slowly without requiring the average carousel rotation speed to change, as long as the time scale for the spark reconfiguration is relatively long.

If, as the above suggests, the sparks reconfigure themselves only slowly after one of the sparks either appears or disappears, one observational consequence of this is that the driftbands should always look connected, which appears to be the case for J0026--1955. This still remains true in the presence of aliasing, although the observed drift rate changes may be magnified. In comparison, \cite{2003A&A...399..223V} point out that connectivity of driftbands is impossible if the carousel rotation rate transitions from non-aliased to aliased regimes.

\subsubsection{Carousel Model in Partially Screened Gap}\label{model2}

Alternatively, a steady change in the carousel rotation rate ($P_4$) with a constant spark configuration is also a plausible explanation for the evolutionary drift modes of J0026--1955. In such a case, some conditions might change at the pulsar surface, making the spark carousel change its rotation speed. As a result, the carousel rotation rate varies smoothly, resulting in a gradual drift rate evolution. Below we provide a hypothesis, using the partially screened gap model \cite[PSG; ][]{2003A&A...407..315G}, for such a phenomenon that can gradually alter the carousel rotation rate.

According to the PSG, a reverse flow of electrons towards the polar cap lead to an ion discharge in the gap region, which ultimately acts as a screen. Due to this screen, the electric field in the polar cap reduces, which reduces the spark velocity \citep[see eqn 28 of ][]{1975ApJ...196...51R}. \cite{2012ApJ...752..155V} also suggests that the dependence of the drift rate on the electric and magnetic fields boils down to a dependence on the variation of the accelerating potential across the polar cap. Hence, a variation in the polar gap results in an increase/decrease in the spark velocity, which in turn may also increase/decrease the carousel rotation rate (regardless of whether the drifting is aliased). Consequently, the carousel rotation rate will also affect $P_3$. In the case of no aliasing, the drift rate will be more negative for a faster carousel rotation rate, and the $P_3$ value will be smaller. Further, as discussed in \cite{2022MNRAS.509.4573J}, a change in drift rate is reflected in the emission heights, where faster drift rate (lower $P_3$ value) emission is thought to originate from higher emission heights; and at lower emission heights, emission from a slower drift rate (higher $P_3$ value) is observed. For curvature radiation, the only way to observe a change in the emission height at one observation frequency is if there is a change in the magnetic field lines. For a changing emission height, the foot points of the magnetic field lines will be closer to or further from the (dipolar) magnetic pole, implying a changing size of the observed spark carousel. For the evolutionary drift modes of J0026--1955, where the drift rate is seen to vary drastically within a drift sequence, a change in emission height (at a fixed observational frequency) would automatically suggest a variable carousel rotation rate for a fixed carousel configuration.

Extrapolating the PSG model, if the screening increases due to the ion discharge in the polar cap region, the electric field in the polar gap region will decrease, lowering the spark velocity and, consequently, lowering the carousel rotation rate. Given the proposed relation between $P_4$ and emission heights, the emission from a carousel with lower $P_4$ will come from a lower altitude in the pulsar magnetosphere. We observe a slow drift rate evolution in mode A1, which could be where the screening steadily increases, causing an evolution towards a slower carousel rotation rate. Similarly, mode A2 can be the case where the screening lowers, causing the drift rate to increase and the emission to come from higher altitudes. Since the typical mode length of observed mode A2 occurrences is not as long as mode A0 or A1, we can only conjecture that the screening cannot lower beyond a certain extent.

For modes A1 and A2, a gradual change in the carousel rotation rate might be an acceptable model. This is also consistent with the general understanding that the rotation rate of the carousel cannot change its magnitude or direction abruptly during a single pulsar rotation (as it implies a rapid change in the pulsar magnetosphere), although it may exhibit a slow evolution.

\subsubsection{Revisiting Mode A2 - Null Correlation}

Our findings suggest a strong correlation between mode A2 and nulling across multiple observation epochs, as discussed in section \ref{sec:drnullcorr}. Such an association indicates the intrinsic nature of these changes and their close relationship with the nulling process. 

The gradual transition from slow to fast drift rate, via addition/reduction of sparks as discussed in section \ref{model1}, might trigger a ``reset'' of the pulsar's radio emission engine, causing the emission to cease for several pulses. J0026--1955 provides compelling evidence for a scenario in which the electromagnetic conditions in the magnetosphere region responsible for radio emission attain a null state after the considerable drift rate evolution seen in mode A2. Additionally, the stability of sparks for fast drift rates could be a reason for short sequences. Implying that for mode A2, the spark configuration becomes unstable once the drift rate is sufficiently fast due to an increase in the number of sparks, causing a null. The instability argument can also explain why mode B sequences only last for a short time compared to all the other modes.

Alternatively, following the subpulse drifting model suggested in section \ref{model2}, mode A2 might indicate the scenario where screening decreases, causing the electric field in the polar gap region to increase. This impacts the spark velocity and, thus, the carousel rotation rate. As the carousel rotation rate increases (due to a decrease in screening), $P_3$ decreases and the driftbands appear closer in the pulse stack. Using the direct dependence between $P_4$ and emission heights (derived from the inverse relation between $P_3$ and emission heights), we can deduce that the emission comes from increasingly higher emission heights for mode A2. Alternatively, it is possible that due to our line of sight, we cannot observe the pulsar past a certain emission height. The drift sequence ends very soon after mode A2 achieves a significantly low drift rate. Similarly, due to the line-of-sight constraint, we might explain why modes with lower drift rates, like mode B, have comparatively shorter mode lengths. 

\section{Summary} \label{sec:conclusion}

We have conducted a thorough analysis of subpulse drifting behaviour in PSR J0026--1955 at 300-500 MHz from uGMRT observations. Our results and conclusions from this study are summarised below.

\begin{enumerate}
    \item From our observations, we have found that the pulsar exhibits short- and long-duration nulls, with an estimated nulling fraction of $\sim58\%$ from our uGMRT observations. The nulling fraction is in stark difference from $\sim77\%$ that was observed at 155 MHz in MWA observations. This disparity could be due to differences in the lengths of observations, or a shallow spectral index component, or a frequency dependence of nulling.
    
    \item The pulsar exhibits unusual drifting behaviour, with both evolutionary and non-evolutionary drift rates. Further, we categorise the unusual subpulse drifting behaviour of this pulsar into two drifting modes: A and B, where mode B is a non-evolutionary mode with a faster drift rate. Mode A was further sub-categorised depending on its evolutionary behaviour. Mode A0 is a non-evolutionary mode with a drift rate 3-4 times slower than mode B. The lack of any curvature in A0 suggests that the viewing geometry must be such that the driftband curvature arising from purely geometric considerations (the ``geometric curvature'') is negligible across the pulse window. Mode A1 is an evolutionary mode which shows a smooth evolution of drift rate from fast to slow. On the other hand, the drift rate in mode A2 evolves from a slower to a faster drift rate. 
    
    \item The individual driftbands for J0026--1955 are not linear and have variable drift rates. We used a cubic smoothing spline estimate on individual driftbands and calculated the gradient (drift rate) at each pulse. To understand the overall drift rate modulation, we fit the drift rates using a quartic polynomial. The model helped in recognising the inter-band and inter-mode variability in drift rates. Though a simplistic higher-order polynomial can describe the global evolution empirically, understanding the local drift rate variability requires more complex modelling. 
    
    \item The pulsar J0026--1955 shows an evolution in drift rate for modes A1 and A2. We advocate the following two models to explain their behaviour:
    \begin{enumerate}
        \item \textit{Variable spark configuration} - The gradual evolution of drift rate in modes A1 and A2 could be caused by a slow change in the spark configuration as the carousel reconfigures into an optimal arrangement after a spark appears or disappears. During this reconfiguration, the angular speed of individual sparks may differ from the average carousel rotation. In this case, the observed evolution in drift rate will be due to the motion of sparks ``under'' the line-of-sight as the entire carousel slowly recomposes.
        \item \textit{Variable carousel rotation rate} -  In this model, we propose that the evolution in drift modes can be explained by a smoothly varying carousel rotation rate with a direct correlation with emission height rather than changes in the number of sparks. The evolution in carousel rotation rate is thought to originate from an increase or decrease in screening in the polar gap region. Therefore, as the screening decreases/increases, the carousel rotates faster/slower (mode A2/A1), with the emission coming from higher/lower altitudes in the pulsar magnetosphere. 
    \end{enumerate} 
    A combination of the two suggested models might also be plausible, a possibility that should be explored in future.
    
    \item J0026--1955 shows robust evidence of subpulse memory across nulls. In multiple instances, we have found the possibility of `drift rate' and `subpulse phase' memory across nulls. We believe that there could be an uninterrupted stable discharge in the polar gap during the null, which is not observed due to the absence of a dominant coherence mechanism or a partially screened gap making the generation of detectable radio emission difficult. 
    
    \item J0026--1955 exhibits an almost consistent behaviour, where a null often follows mode A2. We propose two hypotheses for this behaviour:  
    \begin{enumerate}
        \item The transition from slow to fast drift rates due to the appearance/disappearance of a spark often triggers a ``reset'' of the pulsar’s radio emission engine, which often culminates in a null. This reset is most likely to take place after the occurrence of mode A2, where the pulsar transitions from a slow to a fast drift rate. Not every occurrence of a null sequence is preceded by mode A2. However, most occurrences of mode A2 are followed by a null state. A null, followed by mode A2, must relate to a defined pathway in the pulsar emission as it is consistently observed across all independent data sets (from two epochs of observations). 
        
        \item Using the proposed carousel model, we advocate the idea that a decrease in screening increases the electric field in the polar gap region and impacts the spark velocity and carousel rotation rate. This results in a decrease in $P_3$ and closer driftbands in the pulse stack, suggesting that emission comes from higher emission heights for mode A2. The sequence ends soon after mode A2 achieves a low drift rate, possibly due to our line of sight not being able to observe the pulsar beyond a certain emission height.
    \end{enumerate}
\end{enumerate}

\section*{Acknowledgements}

We thank the anonymous referee for several useful comments that helped improve the paper. PJ acknowledges the Senior Research Fellowship awarded by the Council of Scientific \& Industrial Research, India. We thank S. Kudale for their help with conducting these observations. The GMRT is run by the National Centre for Radio Astrophysics of the Tata Institute of Fundamental Research. 

\section*{Data Availability}

This paper includes data taken from the uGMRT in the 41st observing cycle. 
 



\bibliographystyle{mnras}
\bibliography{references} 








\bsp	
\label{lastpage}
\end{document}